# SpecMon: Modular Black-Box Runtime Monitoring of Security Protocols

## Full Version

Kevin Morio*
kevin.morio@cispa.de
CISPA Helmholtz Center for Information Security
Saarbrücken, Saarland, Germany

Robert Künnemann
robert.kuennemann@cispa.de
CISPA Helmholtz Center for Information Security
Saarbrücken, Saarland, Germany

## ABSTRACT

There exists a verification gap between formal protocol specifications and their actual implementations, which this work aims to bridge via monitoring for compliance to the formal specification.

We instrument the networking and cryptographic library the application uses to obtain a stream of events. This is possible even without source code access. We present an efficient algorithm to match these observations to traces that are valid in the specification model. In contrast to prior work, our algorithm can handle non-determinism and thus, multiple sessions. It also achieves a low overhead, which we demonstrate on the WireGuard reference implementation and a case study from prior work. We find that the reference Tamarin model for WireGuard [20] can be used with little change. We only need to specify wire formats and correct some small inaccuracies that we discovered while conducting the case study. We also provide a soundness result for our algorithm that ensures it accepts only event streams that are valid according to the specification model.

## 1 INTRODUCTION

In today's connected world, security protocols play a vital role in securing communication channels and providing key properties like confidentiality and authenticity. While they are challenging to design, protocol verification tools like ProVerif [11], Tamarin [27], and DeepSec [13] have matured to the point where many protocols can be verified without substantial deviation from their specification, often automatically. These tools operate in the Dolev-Yao (DY) model [19], in which cryptographic messages are modeled symbolically.

While a secure protocol specification is *necessary* for the security of a (faithful) implementation, it is not sufficient. Protocol verification can uncover design-level flaws in abstract models, but not in their implementation. The *verification gap* between the verified specification and the actual application is expensive to close.

The process of implementing protocols is particularly error-prone. Besides general bugs like buffer overflows and off-by-one errors, developers are forced to resolve the underspecified bits of the specification, e.g., which cryptographic algorithm to deploy with which parameters, how to negotiate protocol parameters, and how to use the cryptographic library correctly. Furthermore, great care must be taken to ensure that no secrets are revealed.

Prior works addressed the verification gap with a variety of static analysis techniques. Those provide strong guarantees but are costly, as they require both expert knowledge and considerable effort.

In this work, we propose SpecMon, a lightweight alternative to model checking and theorem proving. Starting from a protocol model given in multiset-rewrite rules (MSRs), the protocol specification language used by Tamarin [27], we derive a *runtime monitor* that ensures observable program events are in accordance to the protocol model. These program events are captured by an *event aggregator* and consumed by the monitor. This approach promises low overhead, flexible integration that is largely independent of the program being monitored, and formal guarantees derived from the verified specification.

Depending on the constraints of the environment—black-box, gray-box, white-box—different techniques are available for event aggregation. Dynamic binary instrumentation toolkits like Frida [17] even allow for black-box monitoring of unmodified binaries. Separating event aggregation from the monitor thus provides us with a very flexible and modular architecture.

We evaluate our approach by conducting two case studies:

- SimpleMAC, a server/client protocol exchanging and verifying HMACs, showcasing after-the-fact instrumentation of an unmodified binary.
- WireGuard, a userspace implementation of the VPN protocol, showcasing monitoring based on annotating Go's cryptographic and networking modules.

The case studies show that SpecMon scales to real-world protocols with acceptable performance and a high degree of automation.

*Contributions.* We make the following contributions in this work:

- Introduce a Tamarin-compatible extension for monitoring by integrating wire formats into MSRs.
- Propose the novel techniques of *rule-splitting* and *trace-rewriting* to connect implementation-level function calls with the symbolic abstraction of the DY model.
- Present an efficient, sound, and complete monitoring algorithm handling non-determinism and multiple sessions.
- Evaluate the performance on the above case studies.
- Examine our changes of the existing WireGuard model.

*Outline.* We discuss related work in Section 2. In Section 3, we detail the architecture of our approach. Section 4 introduces our notation and covers background information on MSRs. The format of protocol specifications is outlined in Section 5, while Section 6 explains the protocol translations. Section 7 introduces our monitoring algorithm, and Section 8 discusses our implementation. The case studies are presented in Section 9, potential limitations in Section 10. We examine future work and conclude in Section 11.

---

*This is an extended version of [29]

## 2 RELATED WORK

While protocol verification traditionally focuses on the specification layer, monitoring is only one of many approaches to ensure that an implementation adheres to a specification satisfying the demanded security properties. Other techniques include code verification (e.g., via separation logic [38]), verified compilers and code generation (cv2ocaml [12], cv2fstar [26], [4]), type checking (F* [39]), model extraction [1, 25, 31], (automated) theorem-proving [24], as well as refinement from specification [9, 33, 7]. All of these are static, which (at least in theory) can ensure security once and for all and thus avoid runtime overhead. The downside is the required expert knowledge and manual effort in applying these techniques.

We take the state-of-the-art analysis by Arquint et al. [7] as an example and motivation, since they also use MSRs as a specification language and evaluate their approach on the same WireGuard implementation as we do. Their end-to-end verification toolchain extracts an input/output (I/O) specification from a Tamarin model and uses off-the-shelf separation logic verifiers to ensure that the source code of the protocol implementation satisfies the extracted specifications. Even though their tool generates a part of the specification, a significant amount of manual work remains. For their WireGuard case studies, 2695 lines (68%) of the specification had to be written by hand.

Pironti and Jürjens [32] proposed an early model driven approach for monitoring security protocol implementations. Starting with the protocol specification in the Spi calculus, they semi-automatically create a monitor implementation that runs in parallel to the protocol implementation. This includes manual refinement steps. Moreover, the Spi calculus only allows for sequential processes, limiting this approach to protocols without parallel execution.

RV-TEE [40] uses trusted execution to monitor cryptographic protocols at runtime, which is exemplified with Elliptic Curve Diffie-Hellman Key Exchange (ECDHKE) in Firefox TLS sessions. Properties are specified as deterministic finite automata. The monitor only supports single TLS sessions. Interleaved TLS sessions must first be separated by a special filtering procedure to determine the start and end of each session. This step is protocol-dependent, requires expertise and should be considered part of the monitoring problem. For performance reasons, the filtering in RV-TEE is based on unsound heuristics, which we would like to avoid, in particular in the security domain. Our approach supports non-determinism and thus multiple sessions, hence it does not require ad hoc filtering. Instead, sessions are distinguished within the mechanism, according to specification, with decent performance and covered by the soundness guarantees.

Runtime verification in the general setting provided inspiration to our algorithm but did not apply directly. Finite automata [8] lack the ability to spawn fresh states and thus deal with an unbounded number of sessions. JavaMOP [23] allows runtime monitoring using different specification languages (e.g., finite automata, LTL, regular expressions), but it is limited to Java programs and faces the same challenges.

Temporal logics like LTL aim to verify properties directly, instead of adherence to a specification that can be verified using, e.g., Tamarin, which we require for two reasons. First, performance: Assuming the specification *can* be verified, the runtime overhead is smaller because verification may run ahead of time. Second, many protocol-specific properties cannot be monitored at all, for instance, if they talk about the attacker's behavior (e.g., secrecy: 'The attacker can never output the secret.') or more than one protocol agent at a time (e.g., agreement: 'The responder accepts a connection for a session only if an initiator requested a matching session.').

The closest match to our approach are Quantified Event Automata (QEA) [8]. Here, event automata are under universal or existential quantification and new instances can be spawned as necessary. Moreover, the algorithm sports excellent performance.

We noticed some key properties about protocol monitoring that render existing approaches for parametric monitoring not (directly) applicable. When monitoring parametric properties, for example, for all events $A(x)$, there exists an event $B(x)$, it is desirable to map multiple instances of $B$ to the same event in the trace. When monitoring a specification, however, each event in the trace must be assigned to one explanation for it in the specification. This uniqueness requirement of the existential quantification directly affects how the monitoring algorithm operates.

Due to these differences in requirements, we developed a new monitoring algorithm for MSRs that greatly benefits from their semantic properties. This had several benefits over an adaptation of QEAs: First, this gives us a compact representation of multiple components and sessions, as the system state is a multiset of facts. Second, outlived state can be automatically pruned, as most facts are linear, limiting the monitor's memory consumption. Third, it simplifies the proof of soundness, as we have a close correspondence between the monitoring algorithm and the semantics of the specification.

There is an unexpected connection between fuzzing and dynamic analysis. Fuzzing can be very useful to find implementation bugs in cryptographic protocols, e.g., in IoT [3]. But systematic attempts at large-scale protocols like TLS require ,besides the usual fuzzing infrastructure, a sizable test suite [37]. The test suite in [37] contains tests for known padding oracle attacks, which is an excellent way of ensuring future versions of a library do not reintroduce vulnerabilities. However, these tests are unlikely to find similar but different attacks. Ammann et al. [6] identify two core challenges to the problem of 'deep fuzzing' protocol implementations. One is to recognize security violations (fuzzing typically recognizes bugs when the program crashes or a buffer overflow occurs). Here, our monitor can help, providing a signal as soon as the implementation deviates from the specification. The second challenge is to reach deep states by producing valid protocol messages. Their solution is a DY-aware fuzzer, which, similar to our method, requires a computable mapping from abstract terms to concrete bitstrings. There are potential synergies concerning the specification of these mappings, which might be explored in future work.

## 3 ARCHITECTURE OVERVIEW

We first describe the architecture of SpecMon on a high level. As depicted in Figure 1, it consists of two independent components—the *event aggregator* and the *runtime monitor*.

The *event aggregator* observes the running implementation and records program events in an event stream. This event stream transcribes the communication between the implementation and its environment. The event stream should be sufficient to monitor the implementation with respect to the specification. Hence, it contains calls to cryptographic and networking functions including their inputs and computed outputs.

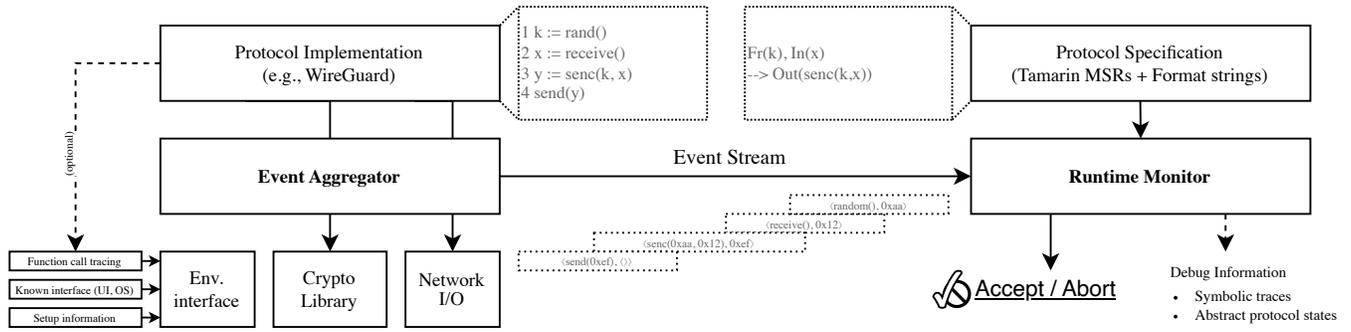

**Figure 1: Architecture overview**

The *runtime monitor* consumes this event stream and decides whether to abort or continue execution. The acceptance requirement is given by a protocol model: If the event stream is a possible instantiation of a valid protocol run of the model, the program may continue, otherwise it is aborted.

Our approach supports *online* and *offline* monitoring. In online mode, the monitor runs in parallel with the implementation. In offline mode, the monitor checks a previously recorded event stream.

### 3.1 Event Aggregator

The primary goal of the event aggregator is to record all *relevant* program events performed by the implementation and create an event stream sufficient for monitoring.

*3.1.1 Aggregation Assumption.* We follow the thesis that the security of a protocol implementation is determined by a sequence of events that the event aggregator fully captures and which can be categorized as follows.

*Cryptographic operations.* Function calls to crypto libraries. We assume the implementation does not implement cryptographic operations by itself. Consequently, our threat model considers the implementation to be benign but possibly faulty. Due to the complexity of these operations, protocol implementations almost exclusively rely on widely established and tested cryptographic libraries.

*Network communication.* Function calls to networking libraries. The assumption is that the implementation uses networking libraries to communicate with the environment and does not try to circumvent monitoring by establishing hidden communication channels.

*Environment interface.* The security properties we want to promote may contain events that require communication between the implementation and its environment. Typical examples include user input or reading files containing keys or configuration data.

In contrast to the former two, the environment interface is implementation-specific. We provide two options for the general case. First, the relevant function, for example, the loading of the

WIREGUARD keys, can be added to the list of monitored functions. Second, the user can specify a script that outputs a setup event, for example by reading a configuration file. The output of the script is then consumed by the monitor before the regular event stream. This solution also provides a high degree of flexibility and works in a black-box setting. It is also possible to include trusted system components (e.g., PIN pads, FIDO-style input devices) or trusted UI elements as part of the environment interface.

The event aggregator needs to capture all program events of these categories with corresponding function symbols in the protocol model to create a complete event stream.

*3.1.2 Syntax and Informal Semantics of Program Events.* A program event has the form $\langle f(x_1, \ldots, x_n), y \rangle$, where $f$ is a function symbol, $x_i$ are the function arguments and $y$ is the return value. If the arguments or return value contain pointers, the event aggregator needs to fetch their values from memory and include them in the program event. This is relevant in particular for cryptographic values, which typically exceed word length. The length of the values are often passed as separate arguments or included in the passed data structure. The resulting event stream is self-contained, such that the monitor can operate without further reference to the system memory. Information that is irrelevant (e.g., status flags in network calls) can be filtered at this stage or later stripped during trace rewriting.

Depending on the programming language and the environment, different techniques can be used to observe the function calls. These include (dynamic) binary instrumentation, source code annotation, and aspect-oriented programming. The event aggregator thus abstracts the idiosyncrasies of the source language and the runtime environment like the calling convention and memory layout.

*3.1.3 Mode of Integration.* The second task of the event aggregator is to integrate with the implementation based on the constraints of the environment.

*Binary instrumentation.* When the source code is unavailable, function calls can be observed using binary instrumentation, leaving the implementation entirely unaffected. This permits debugging in late development stages or monitoring of legacy applications. It requires a symbol table, which compilers like GCC and Clang add by default. In Section 9.1, we use the Frida toolkit [17] to instrument SimpleMAC.

*Annotated libraries.* If the source code is available or the cryptographic library is dynamically linked, we can provide an annotated version. This is the easiest way of integrating the monitor in early and mid development stages and enables continuous tests for compliance or regression. Moreover, the monitor provides easy-to-read debugging information even if the specification model is erroneous or not available at all. For the WireGuard case study (Section 9.2) we created a tool that automatically annotates Go code and use it to annotate Go's crypto and networking modules.

*Verifiable secure closed-source software by isolation.* With suitable isolation or sandboxing techniques, the event aggregator and monitor can be bundled with an application that is dynamically linked against an annotated library. This would allow for shipping closed-source software with guarantees similar to open-source software, while still providing companies with the guarantee that their intellectual property is protected. We did not implement this method, as it requires more tooling, but point to Alrawi et al. [5] and Shu et al. [36] for surveys on various isolation approaches.

*Reusability.* As we typically only need to observe functions that are provided by established libraries and APIs, there is great potential for reusing the event aggregator across different implementations. Since the event aggregator is a separate component, it can be implemented in a language that is optimal for the task. The monitor, however, is always the same.

### 3.2 Runtime Monitor

The runtime monitor checks the event stream recorded by the event aggregator against the specification model. This model describes the protocol behavior as a set of MSRs that encode the protocol's state transitions. It is usually derived from a specification document like an RFC.

The model describes different roles in the protocol such as client and server. The user chooses a role and the framework selects the relevant rules for monitoring. These are further decomposed (cf. Section 6.1) so they perform at most one computation at a time. We call the annotations pertaining to this computation a *trigger*. A rule is applicable if some rule's trigger matches the observed program event and if that rule is applicable in the current configuration. A configuration consists of a multiset of facts representing the current state of the protocol and a trace of rule actions.

In general, it is possible that more than one rule is applicable, and we obtain multiple competing follow-up configurations. This may happen if the protocol's state is not sufficient to uniquely determine the applicable rule. Imagine a multi-session protocol whose messages lack a session identifier. As discussed in Section 2, prior work has tried to avoid non-determinism, whereas we embrace the non-determinism inherent in protocol modeling, for three reasons. Morally, we regard the handling of sessions as an inherent part of the protocol (consider attacks on session handling as in [15, 10]). Theoretically, it increases the trust in the monitor, as the session handling is now covered by its soundness result. Practically, real-life protocols are designed to run efficiently, i.e., messages have identifiers that allow resolving the non-determinism. Otherwise, their implementations would be slow and hard to realize. Indeed, we find our case studies resolve the non-determinism naturally, by themselves, and our monitor never encounters the inefficient non-deterministic case. This should generalize to all protocols that resolve non-determinism by value matching in a way that is visible at the specification (i.e., MSR) level. However, even through non-determinism is often resolved at the protocol level, our MSR-based approach supports non-determinism out of the box, without adding to the complexity of the monitor.

### 3.3 Challenges

The main challenges in implementing the event aggregator lie in interpreting *complex nested data structures* used within the libraries, so that pointers are properly dereferenced, and in *reflecting the library's calling conventions*, e.g., whether computed values are given as return values or written to pointers passed as arguments.

For the runtime monitor, we encountered the following.

*Message formats.* Formal models of security protocols abstract away details that are not relevant for the properties of interest. For one, this concerns the message formats, which we solve by integrating Mödersheim and Katsoris [28] message format into Tamarin' specification language.

*Computation non-determinism.* Protocol models typically leave the order in which computations producing a message take place unspecified. For instance, the term $\langle \mathsf{senc}(k, m), \mathsf{h}(m) \rangle$ could be computed by first performing encryption or by first hashing the message. This produces non-determinism on a different level than the non-determinism already present in the protocol. Fortunately, we can soundly transform the protocol such that we only have a single computation per rule and thus only need to deal with the latter kind of non-determinism in the monitor, simplifying the implementation, the theory, and improving performance. See Section 6.1 for more details.

*Call-sequence abstraction.* More subtly, formal models often simplify how cryptographic functions are actually called. For instance, the term $\mathsf{h}(m)$ can be a series of function calls, at least for large $m$. For example, in Go's `crypto` module `hash.New` initializes the hash computation, `hash.Write` hashes a block of $m$ and `hash.Sum` finalizes the hash computation. The convention for the hash computation is a small protocol in itself, and we indeed solve this problem by reusing the monitor to rewrite the trace in several, optional pre-processing steps. See Section 6.2 for more details.

## 4 BACKGROUND

We introduce the notation we use throughout the paper and recall Multiset-rewriting [27] for the abstract specification of protocols as well as format strings [28] to equip the specifications with more concrete message formats.

*Notation.* The set of natural numbers $\{1, \ldots, k\}$ is denoted by $[k]$, the set of binary numbers $\{0, 1\}$ by $\mathbb{B}$, and the set of bitstrings by $\mathbb{B}^*$. We denote an ordered sequence of elements by $\vec{s}$ and write $\vec{s} \sqcup \vec{t}$ for the concatenation of two sequences. We use $\vec{s}|_X$ for the subsequence of members of $X$ in $s$ and $\vec{s}|_{\neq i}$ for the sequence with the $i$-th element removed. $X^*$ denotes the set of all finite sequences of elements from $X$. We may write $x_i \in \vec{x}$ to denote that $x_i$ is the $i$-th element of $\vec{x}$.

## 4.1 Multiset-rewrite Rules

For protocol specifications we use Tamarin's [27] input language, MSRs. In this language, messages are represented by abstract terms composed of function symbols that represent cryptographic primitives. The abstract semantics of these cryptographic primitives is specified by an equational theory on terms. High-entropy values (e.g., keys or nonces) are constants from an infinite set of names *names*, divided into public (attacker) names $PN$ and secret (protocol) names $FN$. For messages that are expected to come from the network and are thus not yet known, we use a set of variables $\mathcal{V}$. The set of terms $\mathcal{T}(\Sigma_{fact}, PN, FN, \mathcal{V})$, short $\mathcal{T}$, is thus constructed over *names* and $\mathcal{V}$ and applications of function symbols in $\Sigma_{fact}$ on terms. Let $f \in \Sigma_{fact}^n$ denote a function symbol with arity $n$. The set of messages $\mathcal{M}$ is given by the ground terms in $\mathcal{T}$.

To equip terms with a meaning, Tamarin allows the user to define a set of equations $E \subset \mathcal{T} \times \mathcal{T}$. For example, consider

$$E = \{ \text{sdec}(\text{senc}(x, y), y) = x \}$$

to express that decryption reverses encryption with the same key. We define the equivalence relation $=_E$ as the smallest equivalence relation containing $E$ that is closed under the application of function symbols and substitution of variables by terms. We write $\text{sdec}(\text{senc}(m, k), k) =_E m$.

The state of the protocol (typically composed of each role's state) is modeled using a multiset of ground facts $\mathcal{G}$, where a fact $F(t_1, ..., t_k)$ of arity $k$ is ground if all $k$ terms $t_1, ..., t_k$ are ground. There are predefined fact symbols for special purposes. The fact that a name $n \in FN$ is fresh (i.e., was not used before) is marked with the unary fact symbol Fr. Terms being received or sent are represented by In and Out, respectively.

The protocol's behavior is modeled by rewriting rules that operate on this multiset. MSRs are denoted as $l -\![\, a\, ]\!\rightarrow r$. Here, $l$ is called *premise*, $r$ *conclusion* and $a$ *actions*. *Linear* facts used in the premise are consumed by application of the rule. An exclamation mark in front of a fact symbol indicates that it is *persistent* and can be consumed arbitrarily often. For example, freshness Fr is a linear fact, whereas $!A$ is a persistent fact.

EXAMPLE 4.1. $[\, \text{Fr}(k), \text{In}(x)\, ] -\![\, \text{Enc}(x, k)\, ]\!\rightarrow [\, \text{Out}(\text{senc}(x, k))\, ]$ This rule models the encryption of a message $x$ received from the network with a fresh key $k$. The encryption of $x$ is finally sent over the network. We will later use some actions (like $\text{Enc}(x, k)$) as trigger events for the monitor.

Let *lfacts* and *pfacts* denote the linear and persistent facts in a multiset and *set* convert a multiset into a set. Operations on multisets are marked with a $\sharp$ superscript: $\subset^{\sharp}$, $\setminus^{\sharp}$ and $\cup^{\sharp}$.

Let $\mathcal{G}^{\sharp}$ denote a multiset of ground facts and $ginsts_E(\mathcal{R})$ a set of ground instantiations of $\mathcal{R}$ modulo an equational theory $E$. We define a labeled transition relation $\rightarrow_{ginsts_E(\mathcal{R})} \subset \mathcal{G}^{\sharp} \times \mathcal{G}^{\sharp} \times \mathcal{G}^{\sharp}$ as follows.

$$\frac{l -\![\, a\, ]\!\rightarrow r \in ginsts(\mathcal{R})_E \quad lfacts(l) \subset^{\sharp} S \quad pfacts(l) \subset S}{S \xrightarrow{a}_{ginsts_E(\mathcal{R})} (S \setminus^{\sharp} lfacts(l)) \cup^{\sharp} r}$$

$$
\begin{aligned}
[\, \text{Out}(x)\, ] &\quad -\![\,\, ]\!\rightarrow \quad [\, \text{K}(x)\, ] \\
[\, \text{Fr}(x : fresh)\, ] &\quad -\![\,\, ]\!\rightarrow \quad [\, \text{K}(x : fresh)\, ] \\
[\,] &\quad -\![\,\, ]\!\rightarrow \quad [\, \text{K}(x : pub)\, ] \\
[\, \text{K}(x_1), ..., \text{K}(x_k)\, ] &\quad -\![\,\, ]\!\rightarrow \quad [\, \text{K}(f(x_1, ..., x_k))\, ] \\
[\, \text{K}(x)\, ] &-\![\, K(x)\, ]\!\rightarrow [\, \text{In}(x)\, ]
\end{aligned}
$$

**Figure 2: The set of message deduction rules MD.**

A rewriting step is possible if all facts in $l$ are in the current state $S$. In the resulting state, all linear facts from $l$ are removed and all facts in $r$ are added. Using the labelled transition relation, we can define executions and traces of a set of MSRs.

DEFINITION 4.1 (Executions and Traces). *An execution of a set of MSRs is a sequence of transitions from multisets of ground facts labelled by actions starting with the empty multiset*

$$[\,] \xrightarrow{a_1}_{ginsts_E(\mathcal{R})} S_1 \cdots \xrightarrow{a_n}_{ginsts_E(\mathcal{R})} S_n,$$

*such that no fresh name is chosen twice, i.e., for the special fact symbol Fr, we have $S_{i+1} \setminus^{\sharp} S_i = Fr(n)$ and $S_{j+1} \setminus^{\sharp} S_j = Fr(n)$ implies $i = j$. The set of executions is denoted by $execs(\mathcal{R})$, The set of traces $traces(\mathcal{R})$ is the projection on the actions in $execs(\mathcal{R})$. The set of filtered traces without empty actions is given by*

$$traces(\mathcal{R})|_{[\,]} := \{ \langle a_i \rangle_{1 \leqslant i \leqslant m, a_i \neq [\,]} \mid \langle a_1, ..., a_m \rangle \in traces(\mathcal{R}) \}$$

Tamarin combines a user-defined set of rules describing the protocol itself with the built-in rules for message deduction MD depicted in Figure 2. These rules represent a standard DY attacker who obtains knowledge (K) by eavesdropping on the network (Out), creating fresh names, or using public values. The knowledge can be used to construct new terms by applying function symbols. Known terms can be send over the network.

## 4.2 Format Strings

Mödersheim and Katsoris [28] introduce a formal definition and small language of *format strings* to describe and enrich cryptographic messages abstracted by algebraic terms with details of their actual formatting and structure.

They define two properties for a set of formats used by a protocol. First, a format is *unambiguous* if it can be parsed in at most one way. Second, the formats are *pairwise disjoint* if a string can be parsed as at most one format. If both properties are satisfied, the formats can be *soundly* abstracted with free function symbols [28, Theorem 4]. That is, if there is an attack in the model using formats, then there is an attack in the abstract model.

A format definition consists of a function symbol with a body. The body represents a concatenation of format fields. A format field can be a constant or a variable with a length parameter. The length of a variable can either be constant or be a variable that occurred in a previous field of the format. The last field, and only the last field, can omit the length parameter, in which case the variable captures the remainder of the bitstring. This is a restricted set of format fields described by Mödersheim and Katsoris [28] that is still expressive enough to capture common message formats. Fortunately, any format only using these fields is unambiguous by construction [28, Theorem 1].

```
functions:
  payload/3,
  data/2

#ifedef MONITOR
macros:
  payload(t,m,h) = cat(int(l,'8'),byte(t,'1'),string(m,l),byte(h)),
  data(t, m) = cat(int(l, '8'), byte(t, '1'), string(m, l))
#endif

rule Server [role=Server]:
  let  p = payload('0x02', m, h)
       hp = hmac('secret', data('0x02', m)) in
  [ In(p) ] --[ ServerAccept(m), Eq(h, hp) ]-> [ Accept(h, hp) ]
```

```
equations:
  payload(t, m, b) = <t, m, b>,
  data(t, m) = <t, m>
```

**Figure 3: MSR with format strings from SimpleMAC**

Example 4.2. Consider the following format string.

$$msg(m) \coloneqq \mathrm{cat}(\mathrm{byte}(\mathtt{0x01}), \mathrm{byte}(l, 2), \mathrm{byte}(m, l))$$

This format string matches bitstrings starting with the byte $\mathtt{0x01}$, then taking two bytes being interpreted as the length of m. In this way, $msg(\mathtt{0xcafe})$ would match $\mathtt{0x010200cafe}$.

All defined format strings like $msg$ in the above example are abstracted as function symbols contained in the set $\Sigma_{fs}$. By default, we assume integers are big-endian encoded. If a little-endian encoding should be used instead, we can use the reverse operation. In the example above, $\mathrm{byte}(m, \mathrm{reverse}(l))$ would match $\mathtt{0x010002cafe}$.

We extended the original definition such that constants and variables can be typed. In the above example, byte could be replaced by int or string. This is mainly used to for debugging and output formatting. Functionally speaking, the byte type is sufficient since the endianness can be expressed using the reverse operation. Moreover, we added bitwise operations (and, or) and a arithmetic operation (add) to the format string language. These operations may only be used for constructing messages from known values but not for parsing them.

## 5 PROTOCOL SPECIFICATION

We combine Tamarin's MSRs with format strings in a way that allows for using the same model file for monitoring and verification. The following low-level operations are supported in format string definitions by the monitor: int, byte, string, cat, and, or, add with arity two and reverse with arity one.

Due to the soundness result of Mödersheim and Katsoris [28] (cf. Appendix B), we can abstract these operations for verification and only need to add the defining function symbol to the term algebra. To make these transparent for the adversary, the user needs to define accessor functions with the suitable equations or directly equate them to the built-in pair function in Tamarin. The function definitions and equations are only used for verification and ignored by the monitor. The body of the format strings are specified as macros, which are again ignored by Tamarin using its preprocessing feature. In this way, we use the abstract function symbol for verification while the monitor uses the concrete format string for parsing and construction of bitstrings. From the list of operators above, addition is also supported in Tamarin by leveraging the recently added support for natural numbers [15].

Figure 3 gives a concrete example by showing the server role of the SimpleMAC protocol (cf. Section 9.1).

## 6 PROTOCOL TRANSFORMATIONS

To perform monitoring according to the specification given as MSRs, we need to establish a connection between the function symbols used therein and the actual function calls in the implementation. First, we explain how nested function applications in MSRs are decomposed into a set of rules, each rule computing at most one function at a time. Then, we introduce a trace-rewriting mechanism that allows for encoding complex library conventions and bridges the gap between the abstract representation (e.g., $\mathsf{h}(\cdot)$ for hash computation) and a series of library calls (e.g., $\mathtt{hash.New}$, $\mathtt{hash.Write}$, and $\mathtt{hash.Sum}$) as used in the real world.

### 6.1 Rule Decomposition

MSRs specify how messages are constructed and sent, but not the order in which they are computed.

Example 6.1. In the following rule, either $\mathsf{h}(x)$ or $\mathsf{h}(y)$ can be computed first. The order is not specified. However, both have to be computed before f.

$$[\, S_0(x, y) \,] \rightarrow [\, S_1(\mathsf{f}(\mathsf{h}(x), \mathsf{h}(y))) \,]$$

To simplify monitoring, we decompose each MSR so that each decomposed rule contains at most one function symbol (excluding format strings) while preserving the protocol semantics.

Computation is encoded in function symbols, while communication and freshness generation are encoded in special facts (In, Out, and Fr). We first focus on function symbols and ordinary state facts, e.g., $S_0$ and $S_1$ in Example 6.1.

In the previous example, the last computation must be f. We assume we already have a fact $\mathsf{ST}_{\mathsf{h}(x):\cdot}(\langle x, y \rangle, h_x)$ that tells us that the subterm $\mathsf{h}(x)$ evaluated on the variable list $\langle x, y \rangle$ results in the term $h_x$. In the monitor, we use $h_x$ to store the result of the computation of $\mathsf{h}(x)$. Let us also assume that we have a similar fact expressing the computation of $\mathsf{h}(y)$. We can then describe the computation of the term $\mathsf{f}(\mathsf{h}(x), \mathsf{h}(y))$ as follows:

$$\begin{bmatrix} \mathsf{ST}_{\mathsf{h}(x):\cdot}(\langle x, y \rangle, h_x) \\ \mathsf{ST}_{\mathsf{h}(y):\cdot}(\langle x, y \rangle, h_y) \end{bmatrix} \rightarrow [\, \mathsf{ST}_{\mathsf{f}(\mathsf{h}(x), \mathsf{h}(y)):\cdot}(\langle x, y \rangle, f(h_x, h_y)) \,]$$

Subterms that contain no function application are derived from a *start rule*, which contains the original premise.

$$[\, S_0(x, y) \,] \rightarrow [\, \mathsf{ST}_{x:\cdot}(\langle x, y \rangle, x), \mathsf{ST}_{y:\cdot}(\langle x, y \rangle, y) \,]$$

These are used by the *mid rules*.

$$[\, \mathsf{ST}_{x:\cdot}(\langle x, y \rangle, x) \,] \rightarrow [\, \mathsf{ST}_{\mathsf{h}(x):\cdot}(\langle x, y \rangle, \mathsf{h}(x)) \,]$$
$$[\, \mathsf{ST}_{y:\cdot}(\langle x, y \rangle, y) \,] \rightarrow [\, \mathsf{ST}_{\mathsf{h}(y):\cdot}(\langle x, y \rangle, \mathsf{h}(y)) \,]$$

Finally, the *end rule* collects the subterm and produces the original conclusion.

$$[\, \mathsf{ST}_{\mathsf{f}(\mathsf{h}(x), \mathsf{h}(y)):\cdot}(\langle x, y \rangle, t) \,] \rightarrow [\, S_1(t) \,]$$

Observe that only the mid rules contain function applications, but never more than one. Hence, we can annotate them with a trigger event $\mathsf{Trig}(f, \langle x_1, \ldots, x_n \rangle, y)$, where $f$ is the function symbol, $x_i$ a set of argument terms, and $y = f(x_1, \ldots, x_n)$ the result of the function application. In general, it is possible that a subterm $s$ occurs multiple times. Instead of recomputing it, we produce sufficiently many facts $\mathsf{ST}_{s,t}$, which we distinguish with a second index $t$, the parent term of the first argument at the top-level in $s$ occurs at the top-level in a fact. In the end rule above, for instance, we would have the index $\mathsf{ST}_{f(\mathsf{h}(x),\mathsf{h}(y));\top}$, and in the start rule the indexes $\mathsf{ST}_{x;\mathsf{h}(x)}$ and $\mathsf{ST}_{y;\mathsf{h}(y)}$.

We now formalize this procedure. To simplify the presentation, we only consider premises and conclusions with one fact each and assume that actions do not contain any function application. The complete definition with multiple facts is given in Appendix D. Let $fs^n(t)$ denote the set of function symbols at depth $n$ in $t$ and $P: \mathcal{T} \times \mathcal{T} \to 2^{\mathcal{T}} \cup \{\top\}$ return the parent terms of the first argument in the second, i.e., $P(f(\vec{a}), t)$ returns all subterms of $t$ that contain $f(\vec{a})$ as an argument and $\top$ if $f(\vec{a}) = t$. We extend $P$ to sequences of terms by applying it to each term in the sequence.

**Definition 6.1** (Split Rule (simplified)). *Let $r$ be an MSR of the form*

$$\underbrace{[\, \mathsf{F}(v_1, \ldots, v_m)\,]}_{=:\vec{v} \in \mathcal{V}^m} \xrightarrow{e} \underbrace{[\, \mathsf{G}(t_1, \ldots, t_n)\,]}_{=:\vec{t} \in \mathcal{T}^n}.$$

*Then it can be decomposed into*

$$sr(r) := \{\, r^{start} \,\} \cup \{\, r_f^{mid} \mid f \in fs(\vec{t})\,\} \cup \{\, r^{end}\,\},$$

*with*

$$r^{start} := [\, \mathsf{F}(\vec{v})\,] \longrightarrow \left[\, \mathsf{ST}_{x_i; f(\vec{x})}(\vec{v}, x_i) \,\middle|\, \begin{array}{l} x_i \in \vec{x} \\ f(\vec{x}) \in fs^0(\vec{t}) \end{array}\right]$$

$$r_{f(\vec{a})}^{mid} := [\, \mathsf{ST}_{a_i; f(\vec{a})}(\vec{v}, x_{a_i}) \mid a_i \in \vec{a}\,] \xrightarrow{[\, \mathsf{Trig}(f, \vec{x}, f(\vec{x}))\,]}$$
$$[\, \mathsf{ST}_{f; g}(\vec{v}, f(\vec{x})) \mid g \in P(f(\vec{a}), \vec{t})\,]$$

$$r^{end} := [\, \mathsf{ST}_{f; \top}(\vec{v}, x_f) \mid f \in fs^{\infty}(\vec{t})\,] \xrightarrow{e} [\, \mathsf{G}(\vec{t'})\,]$$

*where $x_f \notin vars(r)$ and $t'_i = x_{t_i}$ if $t_i \in fs(\vec{t})$ and $t_i$ otherwise. If $fs(\vec{t}) = \emptyset$, let $sr(r) := r$.*

We lift this step to a set of rules: $SR(\mathcal{R}) := \bigcup_{r \in \mathcal{R}} sr(r)$. Each rule in this set contains at most one function application. Either it did not contain any function application in the beginning ($fs(\vec{t}) = \emptyset$), or the produced rule contains no function application (if it is a start or end rule) or exactly one (if it is a mid rule).

### 6.1.1 Special Facts.

As illustrated in Figure 2, the facts In, Out, and Fr have a special meaning in Tamarin. Therefore, we interpret them as function calls for sending and receiving messages over the network or calls to the cryptographic library's randomness or key generation mechanism. We annotate these rules with the appropriate triggers.[1]

$$[\,] \dashv [\, \mathsf{Trig}(\mathsf{receive}, \langle\rangle, x)\,] \vdash [\, \mathsf{In}(x)\,]$$
$$[\,] \dashv [\, \mathsf{Trig}(\mathsf{random}, \langle\rangle, k)\,] \vdash [\, \mathsf{Fr}(k)\,]$$
$$[\, \mathsf{Out}(x)\,] \dashv [\, \mathsf{Trig}(\mathsf{send}, \langle x\rangle, \langle\rangle)\,] \vdash [\,]$$

Each of these rules has only a single trigger and is thus compatible with the monitor design. The non-determinism is therefore handled by an additional set of calls for the generation of each of the facts In, Out, and Fr. Those can occur in any order.

---
[1]See Section 7.2 and Appendix C for why this is valid.

**Example 6.2.** Consider the MSR

$$[\, \mathsf{In}(m), \mathsf{Fr}(k)\,] \longrightarrow [\, \mathsf{Out}(\mathsf{hmac}(k, m))\,]$$

After receiving a message $m$ (a) and generating a random key $k$ (b), the HMAC of $m$ under $k$ is computed and sent. Steps (a) and (b) can occur in any order.

### 6.1.2 Discussion.

In our approach, the mapping between bitstrings produced by the implementation and terms specified by the model is encoded in the set of ST-facts. Alternatively, the event aggregator could maintain this mapping, e.g., by storing a table of bitstrings and terms. It would then provide the monitor with symbolic traces instead of function calls. Our approach has a significant advantage: It enables us to 'forget' mappings. Once a particular rule application has been unambiguously determined, we can remove facts and the associated bitstrings. If the event aggregator handles the mapping, it would not have access to the protocol's state and thus does not know when to remove the corresponding facts. To avoid running out of memory, we would need to use heuristics such as removing facts by last access.

To benefit from the advantages of MSRs, we chose the first approach. While this requires us to specify when computations *can* take place and when they *must*, it gives us flexibility which terms need to be recomputed. In effect, $SR$ implements a policy for recomputation. Within $sr(\cdot)$, no subterm is computed twice. This is ensured in the mid rule by adding a subterm fact for each $g$ using the respective subterm. Across rules, i.e., in $sr(r_1)$ and $sr(r_2)$, $r_1 \neq r_2$, terms are always recomputed. Note that the order of computation is still arbitrary.

**Example 6.3.** Consider the MSR

$$[\, \mathsf{S}_0(k, x)\,] \longrightarrow [\, \mathsf{S}_1(\mathsf{h}(x), \mathsf{hmac}(k, \mathsf{h}(x)))\,],$$

where $\mathsf{h}(x)$ can be expensive to compute. This is a common idiom: The message is computed and authenticated. An implementation would compute $\mathsf{h}(x)$ once and reuse it for the computation of hmac.

Other evaluation strategies can be encoded prior to decomposition. If $\mathsf{h}(x)$ ought to be reused later, it can be stored in an additional state fact or added to an existing fact. For example,

$$[\, \mathsf{S}_0(x)\,] \longrightarrow [\, \mathsf{Cache}(\mathsf{h}(x)), \mathsf{S}_1(x)\,]$$

Our case studies showed that our default policy is reasonable: $SR$ produces evaluation policies that are flexible and easy to adapt. For SimpleMAC, it matches the order of execution in the implementation, while for WireGuard, due to its concurrency when dealing with multiple peers, a more fine-grained but easy-to-express policy is needed.

### 6.1.3 Lookahead.

Since the start rule does not contain any function symbols, it is not triggered by an observable event. For the monitor to decide which mid rule is triggered and check for its applicability, it must tentatively apply the start rule. This adds artificial non-determinism causing inefficiencies when multiple start rules apply. The monitor cannot directly determine which mid rules are applicable and must try to apply all of them.

This is not a technical artifact, as there is actual non-determinism in the order of computation (Example 6.1). If we tie the start rule (or any other rule starting with $S_1(x, y)$ in the above example) to one of the two possibilities, the same question would arise for the second event (although in our example there is just one possibility), and we would compute all possible interleavings. For efficiency reasons, we need to avoid representing interleavings explicitly, as otherwise, the number of required rules to express them would scale exponentially.

We therefore equip the monitor with a lookahead. The start rule is annotated with a *hint* attribute, which contains the set of events that occur in the mid rules associated with the direct subterms. The semantics, which are implemented in the monitoring algorithm in Section 7, are as follows: A rule with a hint is only considered if the currently observed event matches *any* hint. However, when the rule is applied, the event is not consumed yet. The monitor searches for a mid rule with the same trigger as the hint. If no such rule exists or is applicable, the monitor aborts.

## 6.2 Trace Rewriting

In the DY model, cryptographic values are represented at the level of abstraction that is needed to reason about their security. For instance, a hash function is represented as $h(x)$ because it is handled like an irreversible function symbol, regardless of the internal structure. In an implementation, however, computing a hash is often split into a series of function calls to decrease the required memory. For example, in Go's `crypto` module, a new hash instance h is created with `hash.New`, data is written to it with `h.Write`, and the hash is finalized with `h.Sum`. Describing the hash as $\text{Sum}(\text{Write}(x_2, \text{Write}(x_1, \text{New}())))$ would be tedious. Even worse, if the length of $x_1$ and $x_2$ are variable, this transformation may require the introduction of several rules. In our WireGuard case study, this also affected HMAC and AEAD computations.

We found that the calling conventions of the cryptographic library can be described as MSRs and that we could repurpose the monitoring algorithm for *trace rewriting*. There are thus several independent instances of the monitor running on different *layers*. The first layers perform relatively simple and protocol-agnostic rewriting, feeding the output stream into the next layer. The final layer performs the actual monitoring.

We extend MSRs in a backward-compatible way to describe the rewriting mechanism.

**Definition 6.2** (Extended MSRs). *Let $s = l \multimap [\ a\ ] \mapsto r$ be a MSR whose actions can be partitioned into triggers, hints, and events denoted by $\text{trigger}(a)$, $\text{hints}(a)$, and $\text{events}(a)$. We use the same notation for the respective set of actions in a rule. We write $t|_{Trigs}$, $t|_{Hints}$, and $t|_{Events}$ to denote the trace projected by the respective set. Then $s$ is an eMSR if $lhs(s)$ only contains variables or format strings and $act(s)$ only contains one trigger.*

An eMSR without hints or a trigger is called a $\epsilon$-rule. Additional well-formedness conditions ensuring hints are correct w.r.t. possible rule instantiations are summarized in Definition F.1.

**Definition 6.3** (Trace Rewriting). *A set of eMSRs $\mathcal{R}$ rewrites $t_a$ to $t_b$ if there is a $t \in \text{traces}(\mathcal{R})$ with $t_a = t|_{Trigs}$ and $t_b = t|_{Events}$. If $\text{hints}(a) = \emptyset$, we write $l \multimap [\ a\ ] \mapsto r$ as $l \multimap [\ \text{trigger}(a)/\text{events}(a)\ ] \mapsto r$.*

Observe that the semantics implies that a trace rewriting system has to explicitly 'forward' function calls that are not meant to be rewritten. This is necessary to allow the rewriting steps to reject an event stream for incorrect use of the library, e.g., when the

`Write` function is called without a prior `New`. Hence, function calls that are recorded by the event aggregator but are not relevant for monitoring need to be listed explicitly.

*6.2.1 Example.* In our WireGuard case study, the implementation uses the `blake2s` hash function implemented in Go's `crypto` module. Shortening the verbose function names, focusing only on the 256-bit version and removing the (optional) in-buffer output functionality, the following simple rewrite system is sufficient to handle both abstraction and monitoring for correct use:

$$[\,] \multimap [\ \text{Trig}(\text{New256}, \langle\rangle, d)/\epsilon\ ] \mapsto [\ \text{New}(d)\ ]$$
$$[\ \text{New}(d)\ ] \multimap [\ \text{Trig}(\text{Write}, \langle d, x\rangle, \epsilon)/\epsilon\ ] \mapsto [\ \text{S}(d, x)\ ]$$
$$[\ \text{S}(d, x)\ ] \multimap [\ \text{Trig}(\text{Write}, \langle d, y\rangle, \epsilon)/\epsilon\ ] \mapsto [\ \text{S}(d, \text{byte}(x) \| \text{byte}(y))\ ]$$
$$[\ \text{S}(d, x)\ ] \multimap [\ \text{Trig}(\text{Sum}, \langle d\rangle, h_x)/\text{Trig}(\text{h}, \langle x\rangle, h_x)\ ] \mapsto [\,]$$
$$[\ \text{S}(d, x)\ ] \multimap [\ \text{Trig}(\text{Reset}, \langle d\rangle, \epsilon)/\epsilon\ ] \mapsto [\ \text{New}(d)\ ]$$

The first rule captures the creation of a new digest object $d$, which we track as new. The next two rules handle the initial and all subsequent write steps. Observe the format string in the third rule, which is used to accumulate the complete input string in S's second argument. In the fourth rule, this second argument $x$ is the bitstring that will appear in the abstraction $h(x)$ when the actual rewriting takes place. The fifth rule handles a reset, showing that the calling conventions are similar to a finite automaton (one per object), which can be neatly reflected in MSRs.

Besides complex use cases like this, trace rewriting also solves some trivial but annoying real-world problems. Functions with long names can be abbreviated, as in the rule for the single-call digest:

$$[\,] \multimap [\ \text{Trig}(\text{blake2sSum256}, \langle x\rangle, h_x)/\text{Trig}(\text{h}, \langle x\rangle, h_x)\ ] \mapsto [\,]$$

Likewise, as mentioned above, functions that are not relevant for monitoring can be ignored. For instance, if the implementation uses a KDF for secure password storage outside the protocol.

$$[\,] \multimap [\ \text{Trig}(\text{KDF}, \langle x\rangle, y)/\epsilon\ ] \mapsto [\,]$$

## 7 MONITORING ALGORITHM

The monitoring algorithm operates on the set of decomposed eMSRs and an event stream. The main algorithm ProcessEvent is depicted in Algorithm 1.

The monitoring state consists of a set of active configurations. A configuration is a multiset of ground facts over bitstrings and, in rewriting mode, an output trace. A program event is accepted iff there is at least one configuration that allows processing it. This allows for monitoring non-deterministic behavior.

ProcessEvent processes the program event $e$ with each configuration, removing configurations that do not accept $e$ and updating configurations that do. For each rule, it checks if the rule has hints or a trigger and calls the corresponding handler. Recall that eMSRs cannot have both hints and triggers. Rules with neither are iterated afterwards in HandleTriggers. If the respective handler accepts the event without error, it returns a set of next configurations, which extends the updated configurations. If this set is empty and hence no configuration accepted the event, it was not permissible according to the protocol model, and the algorithm aborts. In rewriting mode, our implementation does not permit more than

---

**Algorithm 1** ProcessEvent

---

**Input:** $\mathcal{R}$: set of eMSRs, $C$: set of configurations, $e$: program event
**Output:** *updated*: set of configurations

1: $updated \leftarrow \emptyset$
2: **for** $c \in C$ **do**
3:     **for** $r \in \mathcal{R}$ **do**
4:         **if** $hints(r) \neq \emptyset$ **then**
5:             $next \leftarrow$ HandleHints$(\mathcal{R}, c, r, e)$
6:         **else if** $trigger(r) \neq \emptyset$ **then**
7:             $next \leftarrow$ HandleTriggers$(\mathcal{R}, c, r, e)$
8:         **if** $next = \perp$ **then**
9:             **return** $\perp$
10:         $updated \leftarrow updated \cup next$
11: **if** $updated = \emptyset$ **then**
12:     **return** $\perp$
13: **return** $updated$

---

**Algorithm 2** HandleTriggers

---

**Input:** $c$: configuration, $r$: eMSR, $e$: program event
**Output:** *updated*: set of configurations

1: $updated \leftarrow \emptyset$
2: **for** $\sigma_r \in$ ConflictSet$(c, lhs(r))$ **do**
3:     $\rho \leftarrow$ mgs$(e, trigger(r)\sigma_r)$
4:     **if** $\rho = \perp$ **then**
5:         **return** $\emptyset$
6:     $d \leftarrow$ ApplyRule$(c, r, \sigma_r \circ \rho)$
7:     $next \leftarrow$ HandleEpsilon$(d, \mathcal{R})$
8:     **if** $|next| > 0$ **then**
9:         $updated \leftarrow updated \cup next$
10:     **else**
11:         $updated \leftarrow updated \cup \{ d \}$
12: **return** $updated$

---

**Algorithm 3** HandleHints

---

**Input:** $\mathcal{R}$: set of eMSRs, $c$: configuration, $r$: eMSR, $e$: program event
**Output:** *updated*: set of configurations

1: $updated \leftarrow \emptyset$
2: **for** $\sigma_r \in$ ConflictSet$(c, lhs(r))$ **do**
3:     $u \leftarrow [\text{mgs}(e, h) \mid h \in hints(r)\sigma_r]$
4:     **if** $|u| \neq 1$ **then**
5:         **return** $\perp$
6:     $\sigma \leftarrow \sigma_r \circ u_0$
7:     $d \leftarrow$ ApplyRule$(c, r, \sigma)$
8:     $D \leftarrow \emptyset$
9:     **for** $s \in \mathcal{R}$ **do**
10:         $next \leftarrow$ HandleTriggers$(d, e, s)$
11:         **if** $next \neq \perp$ **then**
12:             $D \leftarrow D \cup next$
13:     **if** $D = \emptyset$ **then**
14:         **return** $\perp$
15:     $updated \leftarrow updated \cup D$
16: **return** $updated$

---

one configuration, so that only one output trace is produced.[2] We extend the algorithm to process a trace of program events and call it ProcessTrace.

HandleTriggers (Algorithm 2) tries to process an event in a configuration for a rule with a trigger. In line 2, we identify all possible instantiations of rules that match the configuration. For $s, t \in \mathcal{T}$ and $vars(s) = \emptyset$, let mgs$(s, t)$ denote the most general substitution $\sigma$ such that $s = t\sigma$ and $\perp$ if none exists. Since we use this operation for triggers, syntactic unification (i.e., = instead of $=_E$) is sufficient. Similarly, we can define ConflictSet$(c, l)$ as the set of most general substitutions $\sigma$ for which $l\sigma \subset^{\#} c$. Any format string contained in $l$ is evaluated beforehand. ConflictSet gives the set of possible instantiations, which, in line 3, we try to specialize (using mgs) to the trigger. If this is possible (lines 4-6), we apply the rule. Then, we check if an $\epsilon$-rule can be applied and extend the set of updated configurations if possible. It immediately returns if called with $c = \perp$.

Rule Application is performed by the algorithm ApplyRule which, on a configuration $c$, rule $r$ and grounding substitution $\sigma$, returns the updated configuration. If the rule contains a trigger Trig(f, $\langle x_1, \ldots, x_n \rangle, y$), the instantiation is updated to $\sigma' = \sigma \circ [f(x_1, \ldots, x_n) \mapsto y]$. Any format string the rule contains is checked and evaluated. If this or any of the equality constraints Eq$(x, y)$ in $act(r)\sigma'$ is not satisfied, $\perp$ is returned. The updated configuration is given by $(d, t')$ where $d = c \setminus^{\#} lfacts(lhs(r))\sigma' \cup^{\#} rhs(r)\sigma'$ and $t' = t \circ events(r)\sigma'$.

HandleHints (Algorithm 3) is analogous, with the following differences. In contrast to a trigger, rules may contain multiple hints. However, a program event may only match one hint with a unique instantiation (lines 3-5). Then, we iterate over all rules that may be triggered directly by the same event. We need to ensure that each application of a hint rule is followed by a trigger rule whose trigger matches the instantiated hint. We fail if no such rule exists.

HandleEpsilon (Algorithm 4) computes the next configurations for rules without hints or a trigger. It follows the previous two

---

**Algorithm 4** HandleEpsilon

---

**Input:** $\mathcal{R}$: set of eMSRs
      $c$: configuration
**Output:** *updated*: set of configurations

1: $updated \leftarrow \emptyset$
2: **for** $r \in \{ r \mid r \in \mathcal{R}, hints(r) \cup trigger(r) = \emptyset \}$ **do**
3:     **for** $\sigma_r \in$ ConflictSet$(c, lhs(r))$ **do**
4:         $d \leftarrow$ ApplyRule$(c, r, \sigma_r)$
5:         **if** $d \neq \perp$ **then**
6:             $updated \leftarrow updated \cup \{ d \}$
7: **return** $updated$

---

algorithms but does not need to match hints or the trigger. If no rule can be applied, the algorithm returns the empty set.

The implementation performs an indexed lookup for rules instead of iterating over them. Moreover, the implementation actually

---

[2]This restriction could be lifted by spawning multiple instances of subsequent rewriting or monitoring steps, but so far we did not see the need for this.

supports multiple triggers per rule, which is convenient during model development.

## 7.1 Soundness

To be able to map event streams to valid traces in the DY model, they need to match the DY model's assumption of the freshness of keys, i.e., that fresh keys never collide with each other as well as other constants. Hence, for each event stream (i.e., after the fact), we partition the bitstrings into fresh ones (that have been output by the library's RNG) and public ones (that are constants in the protocol). In addition, we require RNG outputs to be unique. We call an event stream with this property *likely*. The precise definition can be found in Appendix E.

With the main algorithm PROCESSTRACE, we can define the monitor output and state the soundness result.

DEFINITION 7.1 (Monitor Output). *For a set of eMSRs $\mathcal{R}$, the output of the monitor is the set of all produced traces*

$$M(\mathcal{R}, t) \coloneqq \{\, t' \mid (t', c) \in \textsc{ProcessTrace}(\mathcal{R}, (\epsilon, []), t)\,\}\,.$$

THEOREM 7.1 (Soundness). *Let $t$ be a likely event stream and $\mathcal{R}$ a set of eMSRs. If $t' \in M(\mathcal{R}, t)$, then there exists an abstraction function $\alpha\colon \mathbb{B}^* \to \mathcal{M}$ and a symbolic trace $t^* \in traces(\mathcal{R})$, such that $\alpha(t) =_E t^*|_{Trigs}$ and $\alpha(t') =_E t^*|_{Events}$.*

The proof can be found in Appendix E.

Intuitively, soundness states that when the monitor accepts an event stream that produces an output trace, this trace can be mapped to a symbolic trace in the specification model.

## 7.2 Soundness in Context

The monitor observes only the behavior of a single role. Hence, the soundness result applies only to a set of rules describing this role's behavior in the protocol. However, security properties about protocols usually concern more than one role. A typical example is authentication, which states that role $A$ accepts a request only if the purported requester $B$ really made that request. We need more than the rules for $A$ to show this property.

Fortunately, Arquint et al. [7] describe syntactic conditions for decomposing a set of MSRs into subsets specific to each role, which they use to statically verify that the implementation of each role adheres to the protocol. These conditions are sufficient to ensure that we produce a model for whichever role $i$ we aim to monitor ($\mathcal{R}^i_{role}$ in their notation). Ensuring these conditions allows us to apply Lemmas 1 and 2 of [7]. For the technical details, see Appendix C.

## 7.3 Completeness

We cannot ensure that the implementation can produce every protocol behavior, but we can ensure that the monitor would permit it. By design, the monitor tracks only one $\epsilon$-transition, which is enough to handle the output of the split rule (Definition 6.1, the end rule) but limits the permitted specifications. This trade-off avoids tracking multiple follow-up states that are reachable via $\epsilon$-transitions and thus increases performance. Prior to the application of $sr$, this only affects rules that contain neither a trigger nor *any* function symbol.

DEFINITION 7.2 (Monitorable). *An eMSR trace $t$ is monitorable if its last element does not contain a hint and if it cannot be extended*

with an action without a trigger, i.e., there is no action $a$ such that $t; a \in traces(\mathcal{R})$ with $trigger(a) = \emptyset$.

THEOREM 7.2 ( Completeness ). *Let $\mathcal{R}$ be a set of eMSRs and $t^* \in traces(\mathcal{R})$ a monitorable trace. Then there exists an abstraction function $\alpha\colon \mathbb{B}^* \to \mathcal{M}$ and $t' \in M(\mathcal{R}, t)$ with $\alpha(t) =_E t^*|_{Trigs}$ and $\alpha(t') =_E t^*|_{Events}$ and $t$ is a likely event stream.*

The proof can be found in Appendix F.

## 8 IMPLEMENTATION

### 8.1 Monitor

We implement the monitor as a standalone program written in the Go programming language. It takes a Tamarin model and the respective agent role to monitor as arguments, parses the model, and decomposes all rules belonging to the specified role as explained in Section 6.1.

Program events are accepted either from standard input (online mode) or from a pre-recorded trace file (offline mode) and the monitor provides the user with a log of the processed events. The user can (optionally) specify the process ID of the monitored program, so the monitor can terminate the process in case of an invalid event. Moreover, the monitor provides a list of permissible events at the time of rejection, which is useful for debugging the implementation or the specification model and even allows for gradually extending underspecified or incomplete models with the missing behavior.

Recall from Section 3 that the environment interface is often protocol-specific. A standard task is the provision of the setup event, typically containing cryptographic keys and other identifiers read from some file. The user may optionally provide a script that produces these events, as an alternative to monitoring such setup functions in the event aggregator.

### 8.2 Event Aggregation

As discussed in Section 3.1, there are various techniques to record program events depending on the programming language and the environment. We exemplify two techniques for event aggregation in Section 9: Frida-based instrumentation for SIMPLEMAC and annotated Go modules for WIREGUARD.

One of our initial goals is to keep the manual work to a minimum. We want to emphasize that both techniques are one-time tasks. The same instrumentation code for Frida can be used to instrument any other program using the same libraries and functions. For Go, the original modules can be easily replaced by our annotated modules using the `replace` directive in the `go.mod` file to enable event aggregation for any Go program using the respective modules.

## 9 CASE STUDIES

We conducted two case studies which are available as a supplementary artifact [30]. First, SIMPLEMAC, an extended version of an example protocol described in Aizatulin et al. [2]. Second, the Go reference implementation of the WIREGUARD VPN protocol [21]. The case studies serve to

(a) evaluate the applicability of our approach,
(b) show its performance characteristics, and
(c) analyze the degree of modification required to make an existing specification compatible with our framework.

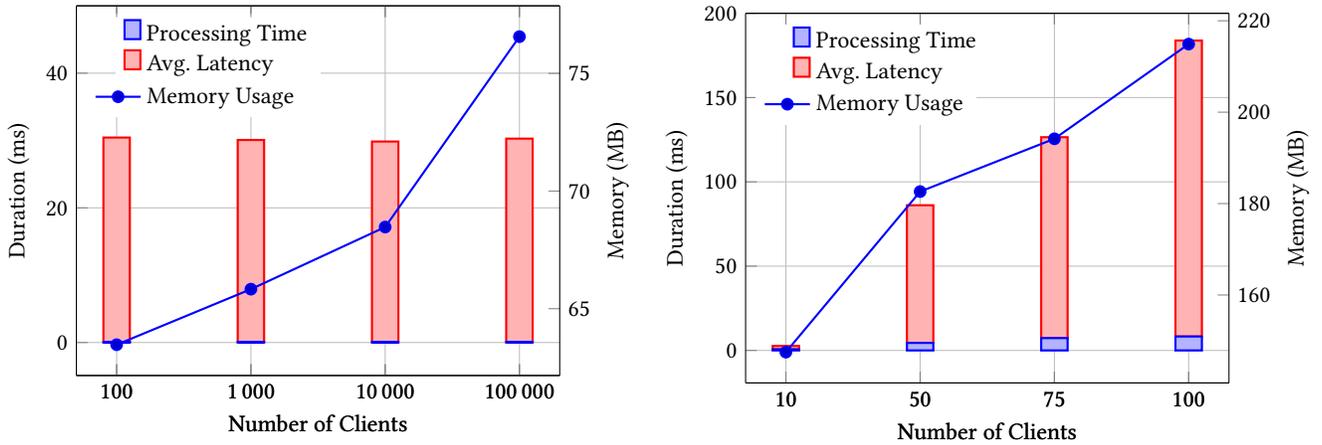

**Figure 4: Evaluation results for SɪᴍᴘʟᴇMAC (left) and WɪʀᴇGᴜᴀʀᴅ (right)**

For the last point, we focus on WɪʀᴇGᴜᴀʀᴅ, as SɪᴍᴘʟᴇMAC was designed to demonstrate the model extraction framework from [2]. We compare our model with the reference Tamarin model [20]. In addition, these case studies serve as a demonstration for two of the three modes of integration discussed in Section 3.1.3.

The results of our evaluation are summarized in Figure 4, visualizing the latency and the combined memory usage of the monitor and event aggregator. The evaluation was performed on an AMD Ryzen Threadripper 2950X and 64 GB of memory.

## 9.1 SɪᴍᴘʟᴇMAC

This case study is based on the HMAC-based message authentication protocol from Aizatulin et al. [2]. Our Tamarin model is derived from a ProVerif model they extracted from their implementation. The function symbols used for encoding and parsing are converted to format strings in the Tamarin model in a direct way. Limitations in their verification method required the original protocol to transmit the payload length in an extra message. We removed this artifact, as incomplete transmission of the payload can already be detected by the HMAC verification. The model consists of three MSRs, one for the generation of the pre-shared key and one for the client and server, respectively. It also includes a security property, which we can verify on the same input file, using Tamarin. It ensures that the message received by the server is the same as the message sent by the client.

We reimplemented the protocol as two standalone programs written in C, one for the client and one for the server, with cryptographic operations are being performed by the OpenSSL library [34]. In contrast to the original implementation of [2], the server is designed to handle multiple clients concurrently.

This case study demonstrates black-box instrumentation (Section 3.1.3). We instrumented three functions using Frida [17], two from the `libc` library (`recv`, `send`) used for network communication and one from the OpenSSL library (`HMAC`) for computing the HMAC. Frida comes with a command-line utility `frida-trace` that can generate stubs for the instrumentation code. Frida's JavaScript API is then used to dereference pointers and extract the function arguments and return values. By using Frida, we can monitor the implementation without modifying its source code nor that of its used libraries. We can thus monitor closed-source software, as long as its symbol table has not been stripped (and even then there are workarounds, e.g., scanning the stack for the argument signature).

We run the case study with one server and scale the number of clients from 100 to 100000 running in parallel. The server is a long-running process that handles all communications with the clients. By increasing the number of clients, we gain insights into the scalability of our approach. Since the size of the configuration depends linearly on the number of clients, we expect the memory usage to increase at least linearly, and possibly worse if the non-determinism is not properly resolved.

Our evaluation results are visualized in Figure 4. We see that our approach achives a very low memory footprint (< 80 MB) even when the number of clients increases. This is due to the automatic state pruning enabled by the MSR semantic and the deterministic nature of the protocol. Due to the latter, the monitor only needs to update a single configuration during the complete run. Moreover, the latency between the arrival of an event at the event aggregator and the time the monitor has processed and accepted the event is low (< 34 ms). As shown in the diagram, this duration is dominated by the latency from the event aggregator until the reception by the monitor. The processing time of each event is below 400 ms.

In addition to the online setting, we performed an offline evaluation using pre-recorded traces. Our monitor is able to process the complete trace of 100000 clients in less than 22 seconds and 24 MB of memory. Hence, our approach is also suitable for post-mortem analysis of faulty implementations.

## 9.2 WɪʀᴇGᴜᴀʀᴅ

We derive a monitor for the WɪʀᴇGᴜᴀʀᴅ implementation from the reference Tamarin model [20]. Ruhault et al. [35] provide a model that is expressed in SAPiC+, which translates into different specification formats, including MSRs. Besides the different specification languages, their model differs mostly in the security properties and corruption scenarios that were considered, which is irrelevant for this work. Not only do we find that the changes are minor and necessary for monitoring (e.g., adding format strings), but they also help us understand where the model is imprecise and where ad hoc abstractions may leave room for incorrect verification results. While we did not find that they are invalidated, imprecise models have resulted in overlooked attacks before (e.g., for the KRACK attack in WPA2 [16, 22]). The monitor is an accessible way to validate models.

**Table 1: Differences compared to [20]**

| Component | Model [20] | This work |
|---|---|---|
| Message format | Tuples | Format strings |
| Long-term keys | Fresh names | Read from files |
| Ephemerals | Precomputed | On demand |
| DDoS protect. | Ignored | Precomputed |
| Counters | Public names | Natural numbers |

Table 1 lists *all* conceptual changes we made to the original model. The first two are inherent. They come with the goal of monitoring implementations. We need to specify the message formats and model that, upon invocation, long-term keys are not freshly drawn but actually read from a file. We modify the corresponding rules such that keys are not bound by Fr-facts but instead by two setup rules with custom triggers, one for initializing the long-term key, the other for initializing each peer's public key. These events are artificially produced by a script that reads the key files produced by wg-genkey and embeds their content into the events. This initial event stream is consumed by the monitor before the actual events are processed. The original model uses pattern matching for verifying the decryption of messages. Pattern matching is not available at the function-call level and can, in general, describe operations that cannot be efficiently computed, such as computing the pre-image of a hash. We did not need to specify roles, because each agent takes both roles simultaneously.

The last three reflect inaccuracies in the reference model [20]. Therein, each type of message is constructed by a single rule. This fixes the order when fresh values are generated and messages are received. In the implementation, ephemeral values are computed independently up to the point when they are first used. Hence, we had to separate the rules for handshake initiation and response to support computation on-demand. Moreover, the concurrency of the implementation required us to further split the rules to reflect possible interleavings and recomputations.

The existing model does not cover WireGuard's DDoS protection messages—presumably because there is no lemma that covers this mechanism. Nevertheless, these constants are precomputed by the implementation even in the case where no DDoS messages are sent, hence we need to include their computation, too. We did not extend the model beyond the missing computation and did not model sending and checking DDoS messages.

Moreover, the prior model abstracts the counters used in the transport messages by using a constant public name. As our monitor supports natural numbers, we changed the model to use these to correctly represent and verify the counter values.

All in all, the required changes are minor. Arguably, they all *increase* the quality of the model. The first two are adding information that is material to the protocol, i.e., how messages are formatted and keys stored, but can be ignored for verification in the DY model. The latter three improve the fidelity and fix inaccuracies of the model.

We can compare this with [7], which uses static analysis to verify WireGuard against a Tamarin specification. First, their Tamarin model is custom, while we based our model on the reference model [20], albeit with modifications. Second, they had to remove code from the implementation of features not included in their Tamarin model (DDoS protection) or otherwise problematic parts (metrics, load balancing). Third, they had to add 3936 lines of specification and proofs (a third generated automatically) to 608 lines of remaining verified code. In contrast, we only instrumented the underlying libraries but did not need to alter the implementation or add any proofs. It is important to emphasize that the changes we had to make to the model only increase its adaptability without imposing any new restrictions. They are hence universal to all WireGuard implementations.

For this case study, we created a tool to automatically annotate Go modules with logging statements. We applied our tool to Go's x/crypto, x/net, and crypto modules, which are used by the WireGuard userspace implementation. The amount of manual work required to instrument the implementation was minimal. It was limited to patching the function for receiving messages over the network. For performance reasons, multiple packets are included in one function call, which we split into individual events. We use trace rewriting to rewrite the function calls into function symbols with bitstrings as explained in Section 6.2.

We run the case study with one instance of WireGuard running as a server and scale the number of client instances from 10 to 100.[3] The server sends a ping message to each client and waits for a response from each client. Finally, the server and all clients terminate.

Our evaluation results are visualized in Figure 4. We see that our approach archives a low memory footprint (< 215 MB) even when the number of clients increases. While the latency between the arrival of an event at the event aggregator and the time the monitor processes it is still acceptable, it increases with the number of clients. This is due to the concurrent nature of the implementation, which generates many program events simultaneously. While the processing time of each event is below 8.3 ms for 100 clients, it is greater than zero and, hence, the latency of each follow-up event increases until all events of the burst have been processed. As before, the latency is also increased due to the encoding and decoding step between the event aggregator and the monitor. To decrease the latency, we could use a more efficient data exchange mechanism like gRPC.

In addition to the online setting, we performed an offline evaluation using pre-recorded traces. Our monitor is able to process the complete trace of 100 clients in less than 15 s and 45 MB of memory. Again, this demonstrates that our approach is also suitable for post-mortem analysis of faulty implementations.

---

[3] The number of clients was limited by the network stack of the operating system.

**Table 2: Summary of assumptions.**

| Assumption | Reference | Intuition and comment |
|---|---|---|
| Aggregation | Section 3.1.1 | Event aggregator collects all relevant interactions. Crucial to our approach but vulnerable to malicious applications. |
| Dolev-Yao | Figure 2 | Cryptography is perfect/adequately captured by the term algebra. Standard in protocol verification. |
| Format Strings | Appendix B | Format strings are disjoint. Has to be checked manually. |
| eMSR | Definition 6.2 | eMSRs contain no patterns (format strings are allowed) and only a single trigger. Ensured when automated rule decomposition is used. |
| Completeness | Definition 7.2 | If any rule before rule decomposition has neither a trigger nor a function symbol, the monitor may be incomplete (but still sound). Such rules are rare. |
| Decomposition | Appendix C | Role-specific rules must access their own state, but they must read it. The state distinguishes sessions. Setup must be done in a specific way. Models that are actually monitorable, i.e., without 'magic' synchronization between parties, can be brought into this form. |

Our evaluation shows that the implementation corresponds to the model. To demonstrate that our approach can also detect deviations from the model, we have deliberately introduced errors into the implementation. In the first instance, we modified the implementation to ensure that the counter in the transport messages is not incremented. In the second case, we altered the implementation to reflect the reuse of ephemeral keys. In both scenarios, the monitor identified the deviation and terminated the implementation.

This case study demonstrates that our approach is also suitable for monitoring complex protocols like WIREGUARD with acceptable performance. The amount of manual work was limited to the manual patching of the network function and the changes to the model. We want to emphasize that both are one-time tasks: Once patched, all programs that use the `x/net` module can be monitored without further modifications, and once changed, the model can be used to monitor different implementations of the WIREGUARD protocol.

## 10 LIMITATIONS

Our approach is flexible and modular and supports a versatile set of monitoring scenarios. Nevertheless, there are limitations that we discuss in this section.

*Model quality.* We require a specification that is realistic enough to be monitored. This was the case for our two case studies, but it is unlikely that the corpus of Tamarin specifications retains this level of quality throughout. Moreover, older models needed to work around features not present in Tamarin (e.g., counters were a recent addition) and thus contain encodings that cannot easily be recognized as such.

*Data-intensive applications.* While our case studies have shown that the monitor incurs only a minor performance and memory overhead, this might be different for throughput-intensive applications like file-sharing applications. Sending large amounts of data may involve a huge number of the respective function calls thereby increasing the memory footprint of both the event aggregator and the monitor.

*Malicious behavior.* We currently target programs with benign errors instead of programs written with malicious intent. Our current approach allows for bypassing the event aggregator by using a statically linked cryptographic library and concealing communication within the payload of the monitored protocol. We encourage research into this problem. Specific protocols allow for recovering security on untrusted systems if there is a trusted *reverse firewall* at the position of the network library [18]. We trust the library and can further inject code into it, as long as it is transparent to the application. Hence, this problem is not harder (and potentially easier) than the reverse firewall problem. Nevertheless, the trust in the cryptographic library remains an inherent limitation albeit orthogonal to our approach. Kleptographic attacks [41], i.e., intentional weaknesses in cryptographic algorithms allowing an attacker to steal information in a way that is undetectable (see the infamous DUAL_EC_DRBG controversy) are outside our threat model. In addition, but less subtly, side channels or other information not captured in the event stream may reveal information.

For an overview of our assumptions, consult Table 2.

## 11 CONCLUSION AND FUTURE WORK

SPECMON provides a flexible, efficient and low-effort method of ensuring an application matches a specification. It can also serve as a building block for other tools, for instance, as an error check in methodologies such as fuzzing [6] or counter-example guided refinement [14]. Another avenue for future research is learning or completing protocol specifications given a coherent set of event streams. The monitor is able to identify when an event cannot be represented in the current state. For simple linear protocols, it may be possible to compute new rules that extend this state. Other cases, like loops or counters, are, of course, more challenging. Nevertheless, obtaining protocol models by linking to an instrumented library seems to be worthwhile.

## REFERENCES

[1] Mihhail Aizatulin, Andrew D. Gordon, and Jan Jürjens. 2012. Computational verification of C protocol implementations by symbolic execution. In *Proceedings of the 2012 ACM Conference on Computer and Communications Security*

(CCS '12). Association for Computing Machinery, Raleigh, North Carolina, USA, (Oct. 16, 2012), 712–723.

[2] Mihhail Aizatulin, Andrew D. Gordon, and Jan Jürjens. 2011. Extracting and verifying cryptographic models from C protocol code by symbolic execution. In *Proceedings of the 18th ACM Conference on Computer and Communications Security* (CCS '11). Association for Computing Machinery, Chicago, Illinois, USA, (Oct. 17, 2011), 331–340.

[3] Fatimah Aljaafari, Lucas C. Cordeiro, and Mustafa A. Mustafa. Verifying Software Vulnerabilities in IoT Cryptographic Protocols. (Jan. 27, 2020). arXiv: 2001.09837 [cs].

[4] José Bacelar Almeida, Manuel Barbosa, Gilles Barthe, and François Dupressoir. 2013. Certified computer-aided cryptography: efficient provably secure machine code from high-level implementations. In *Proceedings of the 2013 ACM SIGSAC Conference on Computer & Communications Security* (CCS '13). Association for Computing Machinery, New York, NY, USA, (Nov. 4, 2013), 1217–1230.

[5] Omar Alrawi, Miuyin Yong Wong, Athanasios Avgetidis, Kevin Valak, and Konstantinos Karak. [n. d.] SoK: An Essential Guide For Using Malware Sandboxes In Security Applications: Challenges, Pitfalls, and Lessons Learned.

[6] Max Ammann, Lucca Hirschi, and Steve Kremer. 2024. DY Fuzzing: Formal Dolev-Yao Models Meet Cryptographic Protocol Fuzz Testing. In 45th IEEE Symposium on Security and Privacy.

[7] Linard Arquint, Felix A. Wolf, Joseph Lallemand, Ralf Sasse, Christoph Sprenger, Sven N. Wiesner, David Basin, and Peter Müller. 2022. Sound Verification of Security Protocols: From Design to Interoperable Implementations (extended version). (Dec. 8, 2022). arXiv: 2212.04171 [cs]. Retrieved Dec. 13, 2022 from http://arxiv.org/abs/2212.04171. preprint.

[8] Howard Barringer, Yliès Falcone, Klaus Havelund, Giles Reger, and David Rydeheard. 2012. Quantified Event Automata: Towards Expressive and Efficient Runtime Monitors. In *FM 2012: Formal Methods*. Vol. 7436. Dimitra Giannakopoulou and Dominique Méry, (Eds.) Red. by David Hutchison et al. Springer Berlin Heidelberg, Berlin, Heidelberg, 68–84.

[9] Karthikeyan Bhargavan, Cedric Fournet, and Andrew D Gordon. 2010. Modular Verification of Security Protocol Code by Typing, 12.

[10] Karthikeyan Bhargavan, Antoine Delignat Lavaud, Cedric Fournet, Alfredo Pironti, and Pierre Yves Strub. 2014. Triple Handshakes and Cookie Cutters: Breaking and Fixing Authentication over TLS. In *2014 IEEE Symposium on Security and Privacy*. 2014 IEEE Symposium on Security and Privacy (SP). IEEE, San Jose, CA, (May 2014), 98–113.

[11] Bruno Blanchet. 2001. An Efficient Cryptographic Protocol Verifier Based on Prolog Rules. In *Computer Security Foundations Workshop*. IEEE Comp. Soc., 82–96.

[12] David Cadé and Bruno Blanchet. 2015. Proved generation of implementations from computationally secure protocol specifications. *J. Comput. Secur.*, 23, 3, 331–402.

[13] Vincent Cheval, Steve Kremer, and Itsaka Rakotonirina. 2018. DEEPSEC: Deciding Equivalence Properties in Security Protocols Theory and Practice. In *2018 IEEE Symposium on Security and Privacy (SP)*. 2018 IEEE Symposium on Security and Privacy (SP). IEEE, San Francisco, CA, (May 2018), 529–546.

[14] Edmund Clarke, Orna Grumberg, Somesh Jha, Yuan Lu, and Helmut Veith. 2000. Counterexample-Guided Abstraction Refinement. In *Computer Aided Verification*. E. Allen Emerson and Aravinda Prasad Sistla, (Eds.) Springer, Berlin, Heidelberg, 154–169.

[15] Cas Cremers, Charlie Jacomme, and Philip Lukert. 2023. Subterm-Based Proof Techniques for Improving the Automation and Scope of Security Protocol Analysis. In *2023 IEEE 36th Computer Security Foundations Symposium (CSF)*. 2023 IEEE 36th Computer Security Foundations Symposium (CSF). (July 2023), 200–213.

[16] Cas Cremers, Benjamin Kiesl, and Niklas Medinger. 2020. A Formal Analysis of {IEEE} 802.11's {WPA2}: Countering the Kracks Caused by Cracking the Counters. In 29th USENIX Security Symposium (USENIX Security 20), 1–17.

[17] Frida Developers. 2024. Frida instrumentation toolkit. (2024).

[18] Yevgeniy Dodis, Ilya Mironov, and Noah Stephens-Davidowitz. 2016. Message Transmission with Reverse Firewalls—Secure Communication on Corrupted Machines. In *Advances in Cryptology – CRYPTO 2016*. Matthew Robshaw and Jonathan Katz, (Eds.) Springer, Berlin, Heidelberg, 341–372.

[19] D. Dolev and A. Yao. 1983. On the security of public key protocols. *IEEE Transactions on Information Theory*, 29, 2, (Mar. 1983), 198–208.

[20] Jason A Donenfeld and Kevin Milner. 2017. Formal verification of the wireguard protocol. *Technical Report, Tech. Rep.*

[21] Jason A. Donenfeld. 2017. WireGuard: Next Generation Kernel Network Tunnel. In *Proceedings 2017 Network and Distributed System Security Symposium*. Network and Distributed System Security Symposium. Internet Society, San Diego, CA.

[22] Changhua He, Mukund Sundararajan, Anupam Datta, Ante Derek, and John C. Mitchell. 2005. A modular correctness proof of IEEE 802.11i and TLS. In *Proceedings of the 12th ACM Conference on Computer and Communications*

[23] Dongyun Jin, Patrick O'Neil Meredith, Choonghwan Lee, and Grigore Roşu. 2012. JavaMOP: Efficient parametric runtime monitoring framework. In *2012 34th International Conference on Software Engineering (ICSE)*. 2012 34th International Conference on Software Engineering (ICSE). (June 2012), 1427–1430.

[24] Jan Jurjens. 2008. Using Interface Specifications for Verifying Crypto-protocol Implementations, 15.

[25] Nadim Kobeissi, Karthikeyan Bhargavan, and Bruno Blanchet. 2017. Automated Verification for Secure Messaging Protocols and Their Implementations: A Symbolic and Computational Approach. In *2017 IEEE European Symposium on Security and Privacy (EuroS&P)*. 2017 IEEE European Symposium on Security and Privacy (EuroS&P). (Apr. 2017), 435–450.

[26] Benjamin Lipp. 2022. Mechanized Cryptographic Proofs of Protocols and their Link with Verified Implementations.

[27] Simon Meier, Benedikt Schmidt, Cas Cremers, and David Basin. 2013. The TAMARIN Prover for the Symbolic Analysis of Security Protocols. In *Computer Aided Verification* (Lecture Notes in Computer Science). Natasha Sharygina and Helmut Veith, (Eds.) Springer, Berlin, Heidelberg, 696–701.

[28] Sebastian Mödersheim and Georgios Katsoris. 2014. A Sound Abstraction of the Parsing Problem. In *2014 IEEE 27th Computer Security Foundations Symposium*. 2014 IEEE 27th Computer Security Foundations Symposium. (July 2014), 259–273.

[29] Kevin Morio and Robert Künnemann. 2024. SpecMon: Modular Black-Box Runtime Monitoring of Security Protocols. In *Proceedings of the 2024 ACM SIGSAC Conference on Computer and Communications Security* (CCS '24). Association for Computing Machinery, Salt Lake City, UT, USA, 15 pages.

[30] [SW] Kevin Morio and Robert Künnemann, SpecMon: Modular Black-Box Runtime Monitoring of Security Protocols (CCS '24 Artifact) Oct. 2024. DOI: 10.5281/zenodo.12787864.

[31] Faezeh Nasrabadi, Robert Künnemann, and Hamed Nemati. 2023. CryptoBap: A Binary Analysis Platform for Cryptographic Protocols. In *Proceedings of the 2023 ACM SIGSAC Conference on Computer and Communications Security* (CCS '23). Association for Computing Machinery, New York, NY, USA, (Nov. 21, 2023), 1362–1376.

[32] Alfredo Pironti and Jan Jürjens. 2010. Formally-Based Black-Box Monitoring of Security Protocols. In *Engineering Secure Software and Systems* (Lecture Notes in Computer Science). Fabio Massacci, Dan Wallach, and Nicola Zannone, (Eds.) Springer, Berlin, Heidelberg, 79–95.

[33] Nadia Polikarpova and Michał Moskal. 2012. Verifying Implementations of Security Protocols by Refinement. In *Verified Software: Theories, Tools, Experiments*. Vol. 7152. Rajeev Joshi, Peter Müller, and Andreas Podelski, (Eds.) Springer Berlin Heidelberg, Berlin, Heidelberg, 50–65.

[34] OpenSSL Project. 2024. OpenSSL. (2024).

[35] Sylvain Ruhault, Pascal Lafourcade, and Dhekra Mahmoud. 2024. A Unified Symbolic Analysis of WireGuard. In *Proceedings 2024 Network and Distributed System Security Symposium*. Network and Distributed System Security Symposium. Internet Society, San Diego, CA, USA.

[36] Rui Shu, Peipei Wang, Sigmund A Zzorzi Iii, Benjamin Andow, Adwait Nadkarni, Luke Deshotels, Jason Gionta, William Enck, and Xiaohui Gu. 2017. A Study of Security Isolation Techniques. *ACM Comput. Surv.*, 49, 3, (Sept. 30, 2017), 1–37.

[37] Juraj Somorovsky. 2016. Systematic Fuzzing and Testing of TLS Libraries. In *Proceedings of the 2016 ACM SIGSAC Conference on Computer and Communications Security* (CCS '16). Association for Computing Machinery, New York, NY, USA, (Oct. 24, 2016), 1492–1504.

[38] Christoph Sprenger, Tobias Klenze, Marco Eilers, Felix A. Wolf, Peter Müller, Martin Clochard, and David Basin. 2020. Igloo: soundly linking compositional refinement and separation logic for distributed system verification. *Proc. ACM Program. Lang.*, 4, (Nov. 13, 2020), 1–31, OOPSLA, (Nov. 13, 2020).

[39] Nikhil Swamy, Juan Chen, Cédric Fournet, Pierre-Yves Strub, Karthikeyan Bhargavan, and Jean Yang. 2011. Secure distributed programming with value-dependent types. In *Proceedings of the 16th ACM SIGPLAN International Conference on Functional Programming* (ICFP '11). Association for Computing Machinery, New York, NY, USA, (Sept. 19, 2011), 266–278.

[40] Mark Vella, Christian Colombo, Robert Abela, and Peter Špaček. 2021. RV-TEE: secure cryptographic protocol execution based on runtime verification. *J Comput Virol Hack Tech*, 17, 3, (Sept. 1, 2021), 229–248.

[41] Adam Young and Moti Yung. 1996. The Dark Side of "Black-Box" Cryptography or: Should We Trust Capstone? In *Advances in Cryptology — CRYPTO '96*. Neal Koblitz, (Ed.) Springer, Berlin, Heidelberg, 89–103.

## A SIMPLEMAC DESCRIPTION

The protocol consists of the roles of a client and a server which agreed on a pre-shared key. The client chooses a message and prepends a tag and the length of the message. It then computes the HMAC of this payload with the pre-shared key. Finally, the client appends the HMAC to the payload and sends it to the server. The server listens for connections of an arbitrary number of clients. When it receives the payload from a client, it splits it into the different components. Since the server knows the pre-shared key, it can recompute the HMAC and compare it to the one received from the client. If the HMACs are equal, the server outputs 1; in the case of a mismatch, it outputs 0.

In the original implementation, the server terminates after one message, because Aizatulin et al. [2] justify their approach using computational soundness, which requires the implementation to have runtime bounds—in their case, a replication bound of 1 per session, i.e., client-server key. Our implementation permits an arbitrary number of sessions but in turn only holds against DY attackers. Moreover, for technical reasons, the size of the initial messages needed to be sent in a separate message, which is very unusual and brittle, as the second message can only be interpreted if the first arrived in time and was picked up by the correct process. As our approach does not have such limitations, we altered the protocol to follow best practices and only send one message. This is sufficient, as an incomplete transmission of the payload can already be detected by the HMAC verification.

## B   SOUNDNESS RESULT OF FORMAT STRINGS

We recall the main result of Mödersheim and Katsoris [28]:

- If we consider a system where all participants except the intruder use only format strings for structuring messages (instead of working with low-level operators like concatenation),
- if all the used format strings are unambiguous and pairwise disjoint, and
- if the system using low-level operators has an attack,
- then there is also an attack in the model that does not use the low-level operators.

**Definition B.1 (Ambiguous Format).** *A format form$(x_1, \ldots, x_n)$ is ambiguous iff there is a string $s$ that can be parsed in two different ways, i.e., as string substitutions $\theta_1$ and $\theta_2$ such that $\theta_1(x_i) \neq \theta_2(x_i)$ for some $1 \leq i \leq n$.*

**Definition B.2 (Disjoint Formats).** *Two formats form$_1(x_1, \ldots, x_n)$ and form$_2(y_1, \ldots, y_m)$ are disjoint iff*

$$[\![\,form_1(x_1, \ldots, x_n)\,]\!] \cap [\![\,form_2(y_1, \ldots, y_m)\,]\!] = \emptyset,$$

*where $[\![\cdot]\!]$ denotes the set of all strings that can be generated by the format.*

## C   ROLE DECOMPOSITION

Rules are annotated with the attribute `role` to bind them to roles in the protocol. To ensure soundness holds within context, the conditions from [7, Appendix A] need to be satisfied:

(1) For all $l \dashv [\, t/a \,] \mapsto r \in \mathcal{R}$:
   (a) $a \in Events$. Our syntax guarantees this.
(2) For all $l \dashv [\, t/a \,] \mapsto r$ that are not bound to a role, i.e., that are environment rules according to the terminology of Arquint et al. [7], the following should hold. This concerns only the rules we fix in Section 6.1.1, or, more precisely, their pre-image w.r.t. the transformation described in [7, Section 3.2.1.]. They represent the interface model, but for presentation reasons, we do not rename the role-specific buffered version for facts In, Out and Fr. This is valid in the execution of the monitor, because it does not interact with any other role's MSRs.
   (a) $l$ and $r$ only use environment facts (namely In, Out and Fr, the reader can verify this) and they are distinct from state facts used in roles. There is no syntactical way to access them, as in the monitor, e.g., In refers to the buffered version, but in Tamarin, In refers to the environment fact before transformation.
   (b) Any rule producing a Setup fact must not produce any other facts on its right-hand side and its label must be empty.
(3) For all $l \dashv [\, t/a \,] \mapsto r$ for a given role:
   (a) $l$ contains only state facts for this role and input facts,
   (b) $r$ only contains state fact for this role and output facts,
   (c) $r$ contains at least one state fact,
   (d) the first argument of each state fact remains unchanged and is of type fresh

For details and the precise notation, we refer to the original work. We combine Lemma 1 and 2 of [7] into a corollary to avoid describing the intermediate model $\mathcal{R}_{\text{intf}}$ [7, Sec. 3.2.1].

**Corollary C.1.** *Let $\mathcal{R}_{\text{role}}^i(rid)$ be the MSR system $\mathcal{R}_{\text{role}}^i$ for a fixed thread id rid. Then*

$$(|||_{i,rid}\mathcal{R}_{\text{role}}^i(rid))||_\Delta \text{role}_{\text{env}}^e \leqslant \mathcal{R}$$

*where $\Delta = [\ldots]$ consists of all ground instances of synchronization labels.*

This follows from $\leqslant\, \subseteq\, \leqslant'$ [7, p. 3], and

**Lemma C.1.** *[7, Lemma 1] $\mathcal{R}_{\text{intf}} \leqslant' \mathcal{R}$.*

**Lemma C.2.** *[7, Lemma 2] Let $\mathcal{R}_{\text{role}}^i(rid)$ be [as above]. Then*

$$(|||_{i,rid}\mathcal{R}_{\text{role}}^i(rid))||_\Delta \text{role}_{\text{env}}^e \leqslant \mathcal{R}_{\text{intf}}, \text{ where [as above.]}$$

To give some intuition, $\mathcal{R}_{\text{role}}^i(rid)$ and role$_{\text{env}}^e$ describe individual LTS for an instance of a role of the protocol and for the environment, i.e., the attacker. For the precise notions of composition ($||_\Delta$ and $|||$) we refer to their work. [38, Corollary 2.4], allows for compositional refinement, i.e., from Theorem 7.1, we can extract what they call a *mediator function* with $\alpha$, and substitute some target $\mathcal{R}_{\text{role}}^x(\cdot)$ by the monitored implementation, provided that we remove all event streams that are not likely (Definition E.1). Let us call the LTS describing the (arbitrary) implementation, intersected with the set of likely traces the monitor accepts $I_{\text{role}}^i(\cdot)$. Overall, assuming that

(1) All other parties are themselves monitored or behave like they are (maybe because they were statically verified, maybe because we trust them), i.e., $I_{\text{role}}^i(rid)$ refines $\mathcal{R}_{\text{role}}^x(\cdot)$ with a mediator function $\Delta_\alpha = \alpha \circ \Delta$.
(2) The real-world environment behaves like a DY attacker, formulated as [7, Proposition 1], which we take as an assumption here: role$_{\text{env}}^e$ refines some concrete environment $\mathcal{E}$ with the same mediator function $\Delta_\alpha$,

we can apply refinement ([38, Corollary 2.4]) to obtain that

$$((|||_{i,rid}I_{\text{role}}^i(rid))||_{\Delta_\alpha}\mathcal{E}$$

This idea is not novel, it follows [7, Section 5.2 and Theorem 2] step by step.

# D SPLIT RULE

We provide the full version of Definition 6.1 and show its correctness.

**Definition D.1** (Split Rule). *Let $r$ be a MSR of the form*

$$[\, \mathsf{F}_1(\vec{v}_1), \ldots, \mathsf{F}_m(\vec{v}_m) \,] \xrightarrow{[\, \mathsf{A}_1(\vec{u}_1), \ldots, \mathsf{A}_\ell(\vec{u}_\ell) \,]} [\, \mathsf{G}_1(\vec{t}_1), \ldots, \mathsf{G}_n(\vec{t}_n) \,],$$

*where $\vec{v}_i \in \mathcal{V}^*$, $\vec{u}_p \in \mathcal{T}^*$, $\vec{t}_j \in \mathcal{T}^*$, $\vec{v} = \bigsqcup_{i=1}^m \vec{v}_i$, and $\vec{t} = \bigsqcup_{j=1}^n \vec{t}_j$
Then it can be decomposed into*

$$sr(r) := \{\, r^{start} \,\} \cup \{\, r_f^{mid} \mid f \in fs(\vec{t}) \,\} \cup \{\, r^{end} \,\},$$

*with*

$$r^{start} := [\, \mathsf{F}_1(\vec{v}_1), \ldots, \mathsf{F}_m(\vec{v}_m) \,] \rightarrow$$
$$\left[ \, \mathsf{ST}_{z_i:f(\vec{t})}(\vec{v}, z_i) \, \middle| \, \begin{matrix} z_i \in \vec{z} \\ f(\vec{z}) \in fs^0(\vec{t}) \end{matrix} \, \right]$$

$$r_f^{mid} := [\, \mathsf{ST}_{a_i:f(\vec{a})}(\vec{v}, x_{a_i}) \mid a_i \in \vec{a} \,] \xrightarrow{[\, \mathsf{Trig}(f, \vec{x}_{\vec{a}}, f(\vec{x}_{\vec{a}})) \,]}$$
$$[\, \mathsf{ST}_{f:g}(\vec{v}_f, f(\vec{x}_{\vec{a}})) \mid g \in P(f(\vec{a}), \vec{t}) \,]$$

$$r^{end} := [\, \mathsf{ST}_{f:\top}(\vec{v}_f, x_f) \mid f \in fs^\infty(\vec{u}) \,] \xrightarrow{[\, \mathsf{A}_1(\vec{u'}_1), \ldots, \mathsf{A}_\ell(\vec{u'}_\ell) \,]}$$
$$[\, \mathsf{G}_1(\vec{t'}_1), \ldots, \mathsf{G}_n(\vec{t'}_n) \,]$$

*where $\vec{v}_f = \vec{v} \sqcup pvars(f)$, $x_f \notin vars(r)$, $t'_{i_j} = x_{t_{i_j}}$ if $t_{i_j} \in fs(\vec{t}_i)$
and $t_{i_j}$ otherwise, as well as $u'_{i_j} = x_{u_{i_j}}$ if $u_{i_j} \in fs(\vec{t}) \cap fs(\vec{t})$ and $u_{i_j}$
otherwise.*

Note that the last point requires that actions may only contain subterms occurring in the conclusion. Each fact ST is implicitly annotated by the rule it originates from. This ensures that each fact ST is only used by instances of the same rule. We may add the following trigger to the produced mid rules $r_f^{mid}$:

$$\mathsf{Trig}(f, \langle a_1, \ldots, a_n \rangle, f(\vec{a})).$$

Observe that the produced rules satisfy the conditions of eMSRs and are thus directly applicable for our monitoring algorithm (cf. Section 7).

**Lemma D.1** (Split Rule Correctness). *Let $SR_r(\mathcal{R}) := (\mathcal{R} \setminus r) \cup sr(r)$.
For all rules $r \in \mathcal{R}$ it holds:*

$$traces(SR_r(\mathcal{R}))|_{[]} = traces(\mathcal{R})|_{[]}.$$

**Proof.** We first show that

$$traces(SR_r(\mathcal{R}))|_{[]} \subseteq traces(\mathcal{R})|_{[]}.$$

Let $\vec{S} \in execs(SR_r(\mathcal{R}))$ be an execution of length $m$ with trace $s$. We show that there exists an execution $\vec{T} \in execs(\mathcal{R})$ of length $n$ with trace $t$ such that

$$S_m \sim_\mathcal{R} T_n$$
$$s|_{[]} = t|_{[]}.$$

The proof is by induction over the number $q$ of end rule transitions in $\vec{S}$. By Lemma D.2, we can assume that $\vec{S}$ is a normal execution.

**Base case** ($q = 0$): If $\vec{S}$ contains no transitions of the end rule, then $\vec{S}$ is an execution of $\mathcal{R}$ by the normal property. The claim follows with $\vec{T} = \vec{S}$ and $t = s$.

**Induction hypothesis** ($q = p$): Let $\vec{S}$ be a normal execution of length $i$ with $p$ end rule transitions and trace $s$. Then there exists an execution $\vec{T} \in execs(\mathcal{R})$ of length $j$ with trace $t$ such that

$$S_i \sim_\mathcal{R} T_j \tag{1}$$
$$s|_{[]} = t|_{[]} \tag{2}$$

**Induction step** ($q = p+1$): Let $\vec{S}$ be a normal execution of length $m$ that contain $p + 1$ end rule transitions. Let $d$ be the index before the $(p + 1)$-th start rule transition. Then $\vec{S}|_{\leqslant d}$ is a normal execution that contains $p$ end rule transitions with trace $s|_{\leqslant d}$. Hence, we can apply the induction hypothesis to get an execution $\vec{T} \in execs(\mathcal{R})$ of length $j$ with trace $t$ such that

$$S_d \sim_\mathcal{R} T_j \tag{3}$$
$$(s|_{\leqslant d})|_{[]} = t|_{[]} \tag{4}$$

Since $\vec{S}$ is a normal execution, $S_d$ is followed by transitions of the start rule, $k$ mid rules, and the end rule. The state after the $k + 2$ transitions is given by

$$S_{d+k+2} = S_d \setminus lhs(r^{start})\sigma \cup [\, \mathsf{G}_1(\vec{t'}_1\sigma), \ldots, \mathsf{G}_n(\vec{t'}_n\sigma) \,]$$

for an instantiation $\sigma$.

Since $lhs(r^{start})\sigma \subseteq S_d$ and $lhs(r^{start}) \subseteq facts(\mathcal{R})$, it follows from (3) that $lhs(r_s)\sigma \subseteq T_j$. Hence, $r\sigma$ is applicable in $T_j$, and we have

$$T_j \xrightarrow{act(r)\sigma} T_{j+1}$$

with

$$T_{j+1} = T_j \setminus lhs(r)\sigma \cup [\, \mathsf{G}_1(\vec{t'}_1\sigma), \ldots, \mathsf{G}_n(\vec{t'}_n\sigma) \,].$$

and trace $t$; $act(r)\sigma$. Hence, with (3) follows $S_{d+k+2} \sim_\mathcal{R} T_{j+1}$. With $act(r)\sigma = act(r^{end})\sigma$ follows

$$(s|_{\leqslant d+k+2})|_{[]} = (s|_{\leqslant d+k+1}; act(r^{end})\sigma)|_{[]}$$
$$= (s|_{\leqslant d}; act(r^{end})\sigma)|_{[]}$$
$$= (t; act(r)\sigma)|_{[]}.$$

If $m = d + k + 2$, then the claim holds. Otherwise, $S_{d+k+2}$ is followed by $m - (d + k + 2)$ transitions of rules in $\mathcal{R}$.

Since $S_{d+k+2} \sim_\mathcal{R} T_{j+1}$, each rule applicable in $S_{d+k+2}$ is also applicable with the same instantiation in $T_{j+1}$. The claim follows.

We next show that

$$traces(\mathcal{R})|_{[]} \subseteq traces(SR_r(\mathcal{R}))|_{[]}.$$

Let $\vec{T} \in execs(\mathcal{R})$ be an execution of length $n$ with trace $t$. We show that there exists an execution $\vec{S} \in execs(SR_r(\mathcal{R}))$ of length $m$ with trace $s$ such that

$$S_m \sim_\mathcal{R} T_n$$
$$s|_{[]} = t|_{[]}.$$

The proof is by induction over the length $n$ of $\vec{T}$.

**Base case** ($n = 1$): Then $\vec{T} = [\, [] \,]$ and $t = \epsilon$ by the MSR semantics. The claim follows with $\vec{S} = \vec{T}$ and $s = t$.

**Induction hypothesis ($n = j$):** Let $\vec{T}$ be an execution of length $j$ with trace $t$. Then there exists an execution $\vec{S} \in execs(SR_r(\mathcal{R}))$ of length $i$ such that

$$S_i \sim_{\mathcal{R}} T_j \tag{5}$$

$$s|_{[\,]} = t|_{[\,]} \tag{6}$$

**Inductive step ($n = j + 1$):** Let $\vec{T}$ be an execution of length $j + 1$ with trace $t$ such that

$$T_j \xrightarrow{act(r^*)\sigma} T_{j+1} \tag{7}$$

for a rule $r^* \in \mathcal{R}$ and instantiation $\sigma$.

There exists an execution $\vec{S} \in execs(SR_r(\mathcal{R}))$ by the induction hypothesis of length $i$ with trace $s$ such that

$$S_i \sim_{\mathcal{R}} T_j \tag{8}$$

$$s|_{[\,]} = (t|_{\leqslant j})|_{[\,]} \tag{9}$$

If $r^* \neq r$, then $r^* \in SR_r(\mathcal{R})$. From (5) and (7) follows $lhs(r^*)\sigma \subseteq S_i$. Hence, $r^*\sigma$ is applicable in $S_i$ resulting in $S_{i+1} = S_i \setminus lhs(r^*)\sigma \cup rhs(r^*)\sigma$ with trace $s' = s; act(r^*)\sigma$. From Definition D.2 and (5) follows

$$\begin{aligned} S_{i+1}|_{\mathcal{R}} &= S_i|_{\mathcal{R}} \setminus lhs(r^*)\sigma \cup rhs(r^*)\sigma \\ &= T_j|_{\mathcal{R}} \setminus lhs(r^*)\sigma \cup rhs(r^*)\sigma \\ &= T_{j+1}|_{\mathcal{R}} \end{aligned}$$

and thus $S_{i+1} \sim_{\mathcal{R}} T_{j+1}$. With (6), we have

$$s'|_{[\,]} = (s; act(r^*)\sigma)|_{[\,]} = (t; act(r^*)\sigma)|_{[\,]} = t'|_{[\,]}$$

Hence, the claim holds.

If $r^* = r$, then $r^{\mathrm{start}}, r_f^{\mathrm{mid}}, r^{\mathrm{end}} \in SR_r(\mathcal{R})$. From (5) and (7) and $lhs(r^{\mathrm{start}}) = lhs(r)$ follows $lhs(r^{\mathrm{start}})\sigma \subseteq S_i$. Hence, $r^{\mathrm{start}}\sigma$ is applicable in $S_i$ resulting in

$$S_i \rightarrow S_{i+1}$$

with

$$S_{i+1} = S_i \setminus lhs(r^{\mathrm{start}})\sigma \cup rhs(r^{\mathrm{start}})$$

Since $ST_{x:f}(\vec{v}\sigma, x\sigma) \subseteq S_{i+1}$, for all $x \in vars(f)$ and $f \in fs^0(\vec{t})$, there exists an instance of the mid rule $r_f^{\mathrm{mid}}$ adding the facts

$$ST_{f:g}(\vec{v_f}\sigma, f(\vec{x}_d)\sigma)$$

for all $g \in P(f(\vec{a}), \vec{t})$. Let $k$ be the number of mid rules. The state after the $k$-th mid rule is given by

$$S_{i+1+k} = S_i \setminus lhs(r^{\mathrm{start}})\sigma \cup \{\, ST_{f:\top}(\vec{v_f}\sigma, f(\vec{x}_d)\sigma) \mid f \in fs^\infty(\vec{t}) \,\}.$$

From the above, we have $lhs(r^{\mathrm{end}})\sigma' \subseteq S_{i+1+k}$, where $\sigma' = \sigma \circ \{x_f \mapsto f(\vec{x}_d)\sigma\}$. Hence, we can apply $r^{\mathrm{end}}\sigma'$ and get

$$S_{i+1+k+1} = S_i \setminus lhs(r^{\mathrm{start}})\sigma \cup [\, G_1(\vec{t'}_1\sigma'), \dots, G_n(\vec{t'}_n\sigma')\,]$$

with trace $(s; act(r^{\mathrm{end}})\sigma')|_{[\,]}$.

Since $x_* \notin vars(r)$ by Definition D.1, we can replace $t_{i_j}$ by $x_{t_{i_j}} = f(\vec{x}_{\vec{a}})$ and $u_{i_j}$ by $x_{u_{i_j}} = f(\vec{x}_{\vec{a}})$ and get[4]

$$\begin{aligned} G_i\sigma &:= G_i(t_{i_1}\sigma, \dots, f(\vec{x}_{\vec{a}})\sigma, \dots, t_{i_q}\sigma), \\ A_i\sigma &:= A_i(u_{i_1}\sigma, \dots, f(\vec{x}_{\vec{a}})\sigma, \dots, u_{i_p}\sigma). \end{aligned} \tag{10}$$

---

[4]For the sake of clarity, we use the same indexes for G and A, even though they are independent.

From (7), follows $T_{j+1} = T_j \setminus lhs(r)\sigma \cup [\, G_1\sigma, \dots, G_n\sigma\,]$ and $act(r)\sigma = [\, A_1\sigma, \dots, A_f\sigma\,]$.

From this, $lhs(r^{\mathrm{start}})\sigma = lhs(r)\sigma$, and (10) follows

$$S_{i+1+k+1} \sim_{\mathcal{R}} T_{j+1}$$

$$(s; act(r^{\mathrm{end}})\sigma)|_{[\,]} = (t|_{\leqslant j}; act(r)\sigma)|_{[\,]} = t|_{[\,]}.$$

Hence, the claim holds. □

**Definition D.2 (State Equivalence Relation).** *Let $S, T \in \mathcal{Q}$ be two states and $F$ be a set of facts.*

$$S \sim_F T \iff S|_F = T|_F$$

*We write $S \sim_{\mathcal{R}} T$ for $S \sim_F T$ if $F = facts(\mathcal{R})$.*

**Definition D.3 (Normal Execution).** *Let $\vec{S}$ be an execution of $SR_r(\mathcal{R})$ with $k$ mid rules. Then $\vec{S}$ is a normal execution if and only if for all transitions $S_i \rightarrow S_{i+1}$ of the start rule, there exists a transition $S_{i+k+1} \xrightarrow{e} S_{i+k+2}$ of the end rule directly preceded by the transitions of the $k$ mid rules.*

**Lemma D.2 (Normal Execution Reordering).** *For all executions $\vec{S} \in SR_r(\mathcal{R})$, there exists an observable-trace equivalent normal execution $\vec{S'} \in SR_r(\mathcal{R})$.*

**Proof.** Let $\vec{S}$ be an execution of $SR_r(\mathcal{R})$ that does not satisfy the normal property.

If $\vec{S}$ does not contain a transition of the end rule, the facts added by any instance of a mid rule, and thus also the start rule, are not consumed by any later rule application. Hence, Corollary D.1 can be used to first remove all transitions of mid rules from last to first and finally the start rule. Since the start and mid rules have an empty action multiset, the observable trace is preserved. The resulting execution only contains applications of rules of $\mathcal{R}$ and thus satisfies the normal property.

Now assume that there exists a transition of the end rule at index $j$. Let $j$ be the smallest such index. By Definition D.1 the end rule is preceded by at least one instance of each mid rule.

Let $d$ be the maximum depth of a term in $\vec{t}$. We first reorder the mid rules $r_f^{\mathrm{mid}}$ with $f \in fs^d(\vec{t})$ such that they follow each other and are finally followed by the end rule. Note that by construction these mid rules are independent of each other and hence their order can be freely chosen. For each $f \in fs^d(\vec{t})$, let $m_f$ be the index of the last transition of $r_f^{\mathrm{mid}}$. Let $\rho_d(f)$ be a fixed but arbitrary ordering of $fs^d(\vec{t})$ and $k_d = |fs^d(\vec{t})|$.

We instantiate Lemma D.3 with $u = m_f$ and $v = j$. We set the offset by which the mid rule at index $m_f$ is shifted to $j - (\rho(f) - k) - m_f - 1 < j - m_f - 1$. In the resulting execution, the shifted mid rules are grouped together and the $k_d$-th mid rule is followed by the end rule. We repeat this step for each $i \in [d]$.

Let $k = |fs(\vec{t})|$. By construction, there exists an instance of the start rule with $\sigma$ at an index $s < j - k - 2$ before any of the mid rules. We can instantiate Lemma D.3 with $u = s$ and $v = j - k - 2$. We set the offset by which the start rule is shifted to $j - k - 3 - s$. In the resulting execution, the start rule is now followed by the $k$ mid rules and the end rule. The transition of the start rule is at index $j - k - 3$.

After the above reordering steps, there may be additional instances of start and mid rules before the shifted start rule at index $j - k - 3$. We shift these rules after the end rule at index $j$. Let $\ell$ be the total number of these rules and let $c_i$ denote the index of the $i$-th such rule. We instantiate Lemma D.3 with $u = c_i$ and $v = j + 1$ for all $i \in [\ell]$ starting from $i = \ell$. We set the offset by which the $i$-th rule is shifted to $j + 1 - c_i$. In the resulting execution the end rule at index $j$ is followed by the $\ell$ rules.

The last reordering step allows us to apply the above approach to next instance of the end rule. Without it, by selecting the largest $m_f$ in the step above, we may select an already reordered mid rule.

After repeating the reordering steps for each end rule, we are left with an execution that may contain start and mid rules without a corresponding instance of an end rule at the end of the execution. With the same argument as in the beginning, we can use Corollary D.1 to remove these spurious transitions. The final execution satisfies the normal property. Since all removed and reordered transitions have an empty action multiset, the observable trace is preserved. □

**Lemma D.3** (Execution Reordering). *Let $\vec{S} \in execs(\mathcal{R})$ be an execution with a transition at index $u$ by rule $r_u\sigma$. Let $w > u$ be such that the transition at index $w$ by rule $r_w\sigma'$ is the earliest transition which requires the transition at index $u$, that is,*

$$lhs(r_w)\sigma' \nsubseteq S_u$$
$$lhs(r_w)\sigma' \subseteq S_{u+1}$$

*If no such transition exists, let $w = |\vec{S}| - 1$. Let $v \leqslant w$.*

*Then, the application of rule $r_u\sigma$ can be shifted to the right from index $u$ up to index $v$ by an offset $k \leqslant v - u$. The resulting execution and trace are given by*

$$\vec{S}' = \begin{cases} S_i & i \leqslant u, i \geqslant u + k + 1 \\ S_{i+1} \setminus rhs(r_u)\sigma \cup lhs(r_u)\sigma & u < i \leqslant u + k \end{cases}$$

$$t' = \begin{cases} t_i & i < u, i > u + k \\ t_{i+1} & u \leqslant i < u + k \\ t_u & i = u + k \end{cases}$$

*with $\vec{S}' \in execs(\mathcal{R})$ and $t \in traces(\mathcal{R})$.*

**Proof.** The reordering is performed by undoing the effects of the application of $r_u\sigma$ at index $u$ for all subsequent states $S_i$ for $u < i \leqslant u + k$. The state $S_{u+1}$ is skipped such that the subsequent states are shifted by one. Since $rhs(r_u)\sigma \subseteq S_i$ for $u < i \leqslant u + k$ by the choice of $v$, the undoing is permissible. By the definition of $\vec{S}'$, we have $lhs(r_u)\sigma \subseteq S_{u+k}$ and thus $r_u\sigma$ is applicable at index $u + k$. The resulting state $S_{u+k+1}$ correctly reflects this application.

For the trace, we note that the above reordering can be seen as moving the action multiset at index $u$ to index $u + k$. The other elements stay unchanged. □

**Corollary D.1** (Rule Removal). *Let $\vec{S} \in execs(\mathcal{R})$ be an execution with trace $t$ and $k \in [|\vec{S}|]$. If for all $i \in [k+1, |\vec{S}|]$, $S_\Delta = S_{k+1} \setminus S_k \subseteq S_i$, then*

$$\vec{S}' = \begin{cases} S_i & i \leqslant k \\ S_{i+1} \setminus S_\Delta & i > k \end{cases}$$

*is an execution of $\mathcal{R}$ with trace $t' = t|_{\neq k}$. Intuitively $\vec{S}'$ is the execution $\vec{S}$ with the $k$-th transition removed.*

**Proof.** By assumption, the facts added by the $k$-th transition are never consumed by a subsequent rule. Hence, Lemma D.3 can be used to shift the $k$-th transition to the end of the execution. Let the resulting execution be $\vec{S}^*$ with trace $t^*$. Then, $\vec{S}' = \vec{S}^*|_{|S^*|-1}$ and $t' = t^*|_{|t^*|-1}$. Since $\vec{S}'$ and $t'$ are prefixes of $\vec{S}^*$ and $t^*$ respectively, they are executions and traces of $\mathcal{R}$ by Definition 4.1. □

# E PROOF OF THEOREM 7.1

Following the DY model, we adopt the assumption that cryptographic functions are correctly implemented, and their computations collide with negligible probability. Moreover, we assume that freshness generation produces distinct values across an execution, and that those not collide with the output of any cryptographic function.

The set of bitstrings is partitioned into two disjoint sets: a set of *fresh* bitstrings, denoted by $\mathbb{B}^{fresh}$, and a set of *public* bitstrings, denoted by $\mathbb{B}^{pub}$. Fresh bitstrings are generated by the RNG, whereas public bitstrings are result of computations and protocol constants.

**Definition E.1** (Likely Event Stream). *An event stream $(t_1, \ldots, t_n)$ is called likely, if the values from the RNG are distinct for each event:*

$$\forall k \in \mathbb{B}^{fresh}. \big|\{ i \mid t_i = \langle \mathtt{random}(), k \rangle \}\big| = 1.$$

From the cryptographic perspective, if there is a non-negligible probability of obtaining an event stream that we call unlikely, then the PRNG must be distinguishable from a mathematically idealized RNG as a distinguisher can both check for PRNG collisions and equivalence with protocol constants.

Finally, we assume that format strings are unambiguous and non-overlapping and hence injective. With these assumptions, we can define an abstraction function relating bitstrings to ground terms. Since program events contain the outer function symbol and a trace of them contains the computations of any subterms, we can reconstruct a binary term representation and map it to its return value.

**Definition E.2** (Bitstring Abstraction). *Let $t$ be an event stream of program events, $t|_{\neq \mathtt{receive}}$ the trace without $\mathtt{receive}$ events and $\mathbb{B}(t)$ the set of all bitstrings in $t$. We define the bitstring abstraction function $\beta_t \colon \mathbb{B}(t|_{\neq \mathtt{receive}}) \to \mathcal{T}(\Sigma_{fact}, \mathbb{B}^{pub}, \mathbb{B}^{fresh})$ such that*

$$\beta_t(y) = \begin{cases} f(\beta_t(x_1), \ldots, \beta_t(x_n)) & if \langle f(x_1, \ldots, x_n), y \rangle \in t \\ k \in \mathbb{B}^{fresh} & if \langle \mathtt{random}(), k \rangle \in t \\ y \in \mathbb{B}^{pub} & otherwise \end{cases}$$

Note that if a bitstring is mapped to itself, then it was either bound by a public variable or by a $\mathtt{random}$ event. We explicitly filter out $\mathtt{receive}$ events in the above definition. When a receive event is observed, there are two possibilities: Either we have seen the bitstring previously and already know a term representation, or it is the first time we see it. In the latter case, its term representation is opaque to us. Since the set of possible ground terms is unrestricted, we can map it to an arbitrary term. When we observe a program event that includes this bitstring as return value, we update the mapping to reflect our gained knowledge.

We can define the symbolic abstraction function. Bitstrings that are returned by program events are mapped to their term representation based on the bitstring abstraction. Public and fresh bitstrings are mapped to public and fresh names respectively. Bitstrings that match a format string are mapped to the format string's function symbol with their arguments recursively mapped to terms. All other bitstrings are mapped to arbitrary terms.

**Definition E.3 (Symbolic Abstraction).** *Let $\omega$ be an injective function mapping $\mathbb{B}^{pub}$ to PN and $\mathbb{B}^{fresh}$ to FN. We extend $\omega$ to function applications homomorphically. Let $\rho$ be an injective function mapping $\mathbb{B}^{pub}$ to $\mathcal{M}$. We define the symbolic abstraction function over a trace of program events $t$ as*

$$\alpha_t(y) := \begin{cases} \omega(\beta_t(y)) & \text{if } y \in dom(\beta_t) \\ fs(\alpha_t(k_1), \dots, \alpha_t(k_n)) & \text{if } match(y, fs), fs \in \Sigma_{fs} \\ \rho(y) & \text{otherwise} \end{cases}$$

**Corollary E.1.** *The function $\alpha_t$ is injective module an equational theory E.*

**Proof.** In the first case, the composition of the two injective functions $\omega$ and $\beta_t$ is again injective. In the second case, by assumption each bitstring matches at most one format string. Finally, $\rho$ is also injective. □

We note that the above definition accounts for new knowledge about the term representation of a bitstring. On first sight, it is mapped to an arbitrary term. Once encountered in a program event, it is mapped to its now known term representation.

We also recall the following requirement (called MSR3) from Tamarin [27], which ensures that the right-hand side of an MSR is always bound by the left-hand side and action. For each rule instance $l \dashv a \rightarrow r \in ginsts_E(\mathcal{R})$, $(\bigcap_{r'=_E r} names(r') \cap FN)$ is a subset of the fresh names in $l$ and $a$.

Rules may contain equality constraints that are checked by the monitor during execution on the bitstring level. However, this does in general not imply term equality. We follow the reasoning of Arquint et al. [7] and assume that bitstring equality implies term equality for two concrete bitstrings (e.g., $x = y \implies \alpha_t(x) =_E \alpha_t(y)$).

**Theorem 7.1 (Soundness).** *Let $t$ be a likely event stream and $\mathcal{R}$ a set of eMSRs. If $t' \in M(\mathcal{R}, t)$, then there exists an abstraction function $\alpha : \mathbb{B}^* \to \mathcal{M}$ and a symbolic trace $t^* \in traces(\mathcal{R})$, such that $\alpha(t) =_E t^*|_{Trigs}$ and $\alpha(t') =_E t^*|_{Events}$.*

**Proof.** We prove the stronger statement that the abstraction function $\alpha_t$ maps $c \in C$ to the last multiset $S$ of a symbolic execution and the projected traces fulfill the theorem statement.

The proof is by induction over the length of $t$. The base case is trivial. In the inductive step, for some $n$, we fix the configuration $c$ and monitored trace $t$ as well as the symbolic multiset $S$ and trace $t'$ for which the above statement holds. In particular, $S = \alpha_t(c)$.

Let $e$ be the next program event. Assume $t; e$ is accepted. The opposite case is trivial. By the induction hypothesis, ProcessEvent accepts $t$ and hence $c \in C$. If $e$ is accepted, either HandleHints or HandleTriggers have returned a next configuration with inputs $\mathcal{R}, c, r, e$ for a rule $r \in \mathcal{R}$.

If $e$ is accepted by HandleTriggers, then there exists at least one $\sigma_r$ such that line 9 or 11 are reached. Then $\sigma = \sigma_r \circ \rho$ with $lhs(r)\sigma \subset^{\#} c$ by definition of ConflictSet. Moreover, $trigger(r)\sigma = e$ by line 3. From this and $S = \alpha_t(c)$ we conclude $lhs(r)(\sigma \circ \alpha_t) \subset^{\#} S$,

Since $e$ is accepted, ApplyRule returns the updated configuration $(d, t_d)$ with $d = c \setminus^{\#} lfacts(lhs(r))\sigma \cup^{\#} rhs(r)\sigma$ and $t_d = t'; (act(r)|_{Events})\sigma$. Moreover, all equality constraints are satisfied. We thus have $(act(r)|_{Trigs}\sigma = e$.

We next discuss how the abstraction function $\alpha_{t;e}$ changes. Assume first that $e$ is a receive event. Then there are two possibilities: Either we encounter the value for the first time or we have seen it before. In the former case, $\alpha_{t;e}$ adds a mapping from the received bitstring to an arbitrary term while keeping the rest of the mapping unchanged. In the latter case, the bitstring is mapped to its already-known term representation. Either case is permissible as there are no constraints for the receive rule. Assume next, that $e$ is a random event. Then the bitstring is mapped to a new fresh name by $\omega$. If $e$ is a send event, then the bitstring being sent is already mapped to its known representation. As neither of these rules contain equality constraints or format strings, they are directly applicable.

Finally, assume that $e$ is any other program event. As before, there are two possibilities: If we see the bitstring for the first time, then it is mapped to its recursively defined term representation. If any of the argument bitstrings are also new and do not occur in $lhs(r)\sigma$, they must have been bound by public variables and are thus correctly mapped to public names by $\omega$. If we have already seen the bitstring, it is mapped to its already-known term representation. Note that by our assumption no two non-receive events can have the same return value. Hence, previous mappings are not overwritten. Next, we need to ensure that format strings are correctly mapped. Assume that some fact in $lhs(r)$ contains a format string $fs$. By the definition of ConflictSet, we know that there exists a fact in $c$ with a bitstring $x$ matching $fs$, i.e., $match(x, fs)$. Recall that a bitstring can either be parsed by a format string or be the return value of a program event. Hence, $\alpha_{t;e}$ correctly maps $x$ to $fs(\alpha_{t;e}(k_1), \dots, \alpha_{t;e}(k_n))$, the term representation of the format string. If any of the $k_i$ are new, they are mapped to arbitrary terms. Note that format strings contained in $rhs(r)$ do not restrict the applicability of $r$ as they are constructed from known values. Next, we need to ensure that the equality constraints are satisfied. Since ApplyRule succeeded, we know by its definition that $\text{Eq}(x\sigma, y\sigma)$ is satisfied for all constraints in $r$. Hence, for all such $x$s and $y$s, we have $x\sigma = y\sigma$ and by assumption $\alpha_{t;e}(x\sigma) =_E \alpha_{t;e}(y\sigma)$.

The above shows that a mapping is only updated if it was previously mapped to an arbitrary term. Mapping it to a more concrete term does not violate any format string or equality constraints. Hence, $lhs(r)(\sigma \circ \alpha_{t;e}) \subset^{\#} S$ and the rule $r$ is applicable in $S$ with transition

$$\alpha_t(c) = S \xrightarrow{act(r)(\sigma \circ \alpha_{t;e})} r(\sigma \circ \alpha_{t;e}) \ S' = \alpha_{t;e}(d)$$

where $S' = S \setminus^{\#} lfacts(lhs(r))(\sigma \circ \alpha_{t;e})) \cup^{\#} rhs(r)(\sigma \circ \alpha_{t;e})$.

If HandleEpsilon returns $\emptyset$, then the updated set of configurations contains $(t_d, d)$ and the theorem statement holds with

$$\alpha_{t;e}(t;e) = (t^*; act(r)\sigma \circ \alpha_{t;e})|_{Trigs}$$
$$\alpha_{t;e}(t_d) = (t^*; act(r)\sigma \circ \alpha_{t;e})|_{Events}$$

If HandleEpsilon returns at least one follow-up configuration, then the algorithms ConflictSet and ApplyRule were called with the same arguments as before, but for rules that contain neither hints nor triggers. For any such rule $r'$ with $d' \neq \bot$ at line 5 of HandleEpsilon, we have

$$d' = d \setminus^{\#} lfacts(lhs(r'))\sigma_{r'} \cup^{\#} rhs(r')\sigma_{r'}$$

$$t'_d = t_d; (act(r')|_{Events})\sigma_{r'}$$

With the same arguments as before follows

$$S' \xrightarrow{act(r')(\sigma_{r'} \circ \alpha_{t;e})} r'(\sigma_{r'} \circ \alpha_{t;e})} S''$$

with $S'' = S' \setminus^{\#} lfacts(lhs(r'))(\sigma_{r'} \circ \alpha_{t;e}) \cup^{\#} rhs(r')(\sigma_{r'} \circ \alpha_{t;e})$ and $S'' = \alpha_{t;e}(d')$. Note that since $r'$ does not contain triggers $\alpha_{t;e}$ as well as the trace of program events stays the same. The trace of events from the previous transition is extended by $act(r')(\sigma_{r'} \circ \alpha_{t;e})$ and thus the theorem statement holds.

If $e$ is accepted by HandleHints, then there exists at least one $\sigma$ such that line 15 is reached. With the same argument as for HandleTriggers we conclude $lhs(r)(\sigma \circ \alpha_t) \subset^{\#} S$. Hence, the rule $r$ is applicable in $S$ with transition

$$\alpha_t(c) = S \xrightarrow{act(r)(\sigma \circ \alpha_{t;e})} r(\sigma \circ \alpha_{t;e}) S' = \alpha_{t;e}(d)$$

where $S' = S \setminus^{\#} lfacts(lhs(r))(\sigma \circ \alpha_{t;e}) \cup^{\#} rhs(r)(\sigma \circ \alpha_{t;e})$. As before, the invariant holds with this transition. HandleHints then calls HandleTriggers on $d$, the instantiated program event $e\sigma$ and any rule $s \in \mathcal{R}$. Since the theorem statement holds for $d$, we can instantiate the above proof for the HandleTriggers case again to show that the theorem statement also holds in this case. □

## F PROOF OF THEOREM 7.2

Definition F.1 (Well-formed Protocol Specification). *A set of eMSR $\mathcal{R}$ is well-formed w.r.t. an equational theory $E$, iff:*

(1) *Hints are exclusive: No state $S$ is reachable with $\rightarrow_{ginsts_E(\mathcal{R})}$ where there are two distinct $r_1, r_2 \in \mathcal{R}$ such that some $r'_t =_E r_i$ and $\sigma$ apply to $S$.*

(2) *Hints are correct: All follow-up rules are reflected in the hint set. If $S \xrightarrow{a}_{ginsts_E(\mathcal{R})} S'$ with $h \in hints(a)$, then for all*

$$S' \xrightarrow{a'}_{ginsts_E(\mathcal{R})} S''$$

*we have $h \in trigger(a')$.*

(3) *$\epsilon$-rules apply once: No state is reachable with $\rightarrow_{ginsts_E(\mathcal{R})}$ where there are two follow-up states*

$$S \xrightarrow{a}_{ginsts_E(\mathcal{R})} S' \xrightarrow{a'}_{ginsts_E(\mathcal{R})} S''$$

*and $trigger(a) = trigger(a') = hints(a) = hints(a') = \emptyset$.*

(4) *Hints and triggers contain all public variables not in the LHS: For all $p \in trigger(r) \cup hints(r)$,*

$$pvars(p) = pvars(r) \setminus pvars(lhs(r)).$$

For completeness, we need term equality to imply bitstring equality. Since $\alpha_t$ is injective, we already have syntactic equality, i.e., $\alpha_t(x) = \alpha_t(y) \implies x = y$. To lift this to equality modulo $E$, we require that for all equations $(t = t') \in E$ and all grounding substitutions $\sigma$, $\alpha_t^{-1}(t\sigma) = \alpha_t^{-1}(t'\sigma)$. We argue that this is s sensible requirement which holds for most common equations, e.g., $sdec(senc(x, y), y) = x$. We can state the completeness of the monitor under the above conditions.

Theorem 7.2 (Completeness). *Let $\mathcal{R}$ be a set of eMSRs and $t^* \in traces(\mathcal{R})$ a monitorable trace. Then there exists an abstraction function $\alpha \colon \mathbb{B}^* \rightarrow \mathcal{M}$ and $t' \in M(\mathcal{R}, t)$ with $\alpha(t) =_E t^*|_{Trigs}$ and $\alpha(t') =_E t^*|_{Events}$ and $t$ is a likely event stream.*

Proof. We proof the stronger statement that:

- The abstraction function $\alpha_t$ maps each fact set $S$ in the execution producing $t^*$ to a configuration labeled $c$ or $d$ in one of the algorithms ProcessEvent, HandleTriggers, HandleHints, or HandleEpsilon.
- These configurations are added to the set of configurations unless further $\epsilon$-transitions are possible, or they contain a hint.
- The projected traces from this execution fulfills the theorem statement.

The proof is by induction on the length of the execution producing $t^*$. The base case is trivial. In the inductive step, for some $n$, we fix the symbolic multiset $S$ and symbolic trace $t^*$ as well as the corresponding configuration $c \in C$ for which the above statement holds. In particular, $S = \alpha_t(c)$.

We assume there exists a transition $S \xrightarrow{a\sigma}_{r\sigma} S'$ for some $r \in \mathcal{R}$. The opposite case is trivial. From this follows

- $S' = S \setminus^{\#} lfacts(lhs(r))\sigma \cup^{\#} rhs(r)\sigma$,
- $lfacts(lhs(r))\sigma \subset^{\#} \alpha_t(c)$,
- $pfacts(lhs(r))\sigma \subset \alpha_t(c)$
- $x =_E y$ for all Eq$(x, y) \in act(r)\sigma$

Let $t^{**} = t^*; act(r)\sigma$. As $\alpha$ is injective, we have

- $lfacts(lhs(r))(\alpha^{-1} \circ \sigma) \subset^{\#} c$,
- $pfacts(lhs(r))(\alpha^{-1} \circ \sigma) \subset c$

From Definition 6.2, $r$ has (1) a trigger, (2) a set of hints, or (3) neither.

*Case (1).* In this case, ProcessEvent or HandleHints call HandleTriggers. The latter then finds $\sigma' = \sigma_r \circ \rho = (\alpha^{-1} \circ \sigma)$ from the definition of ConflictSet and the use of mgs in line 3.

From the above we know that all symbolic equality constraints are satisfied. By our assumption this implies that the equality constraints on the bitstrings are also satisfied. Hence, ApplyRule succeeds and produces $d = c \setminus^{\#} lfacts(lhs(r)\sigma') \cup^{\#} rhs(r)\sigma'$ which is then added to the set of configurations, either explicitly in line 11, or implicitly when running HandleEpsilon on it. The observed event $e$ equals $trigger(r)\sigma'$ and thus $\alpha_t(e) =_E (a\sigma)|_{Trigs}$. ApplyRule extends the current trace $t'$ to $t''$ with $events(r)\sigma'$. Moreover, from $\alpha_{t;e}(events(r)\sigma') =_E events(r)\sigma$ follows $\alpha_{t;e}(t'') =_E \alpha_{t;e}(t^{**});)|_{Events}$ Note that by construction, if $y$ is the result of the computation in $e$, and $\alpha_{t;e}(y) = m$ for some term $m$, then $m \in trigger(r)\sigma$.

The check in line 8 ensures that $d$ is only added to the configuration if no $\epsilon$-transition is possible.

*Case (2).* In this case, ProcessEvent calls HandleHints. The check on line 4 ensuring that there exists only one hint matching the event $e$ is guaranteed by Definition F.1. The rest is analogous to Case (1). Lines 10 and 11 ensure that $d$ is never directly added to the set of configurations. Only the follow-up configurations returned by HandleTriggers are added.

*Case (3).* In this case, HandleTriggers calls HandleEpsilon since, from the induction hypothesis, we never have $c$ in the set of configurations as an $\epsilon$ transition is still applicable. From Definition F.1, we have that $S$ was not reached via an $\epsilon$-transition. The argument for the state invariant is analogous to the argument for Case (1). Again from Definition F.1(3), we conclude that no further $\epsilon$-transitions are possible. Hence, it is correct to add the resulting $d$ to the set of configurations in line 11 of HandleTriggers. □